\documentclass[graybox]{svmult}

\usepackage{mathptmx}       
\usepackage{helvet}         
\usepackage{courier}        
\usepackage{type1cm}        
\usepackage{makeidx}         
\usepackage{graphicx}        
\usepackage{multicol}        
\usepackage[bottom]{footmisc}

\usepackage{amsmath}
\usepackage{bm}

\makeindex             

\begin{document}

\title*{Quark Matter in a Strong Magnetic Background}
\author{Raoul Gatto and Marco Ruggieri}
\institute{Raoul Gatto \at Departement de Physique Theorique,
 Universite de Geneve, CH-1211 Geneve 4, Switzerland. \email{raoul.gatto@unige.ch}
\and Marco Ruggieri \at Department of Astronomy and Physics, Catania University, 
Via S. Sofia 64, I-95125 Catania. \email{marco.ruggieri@lns.infn.it}
}
%
%
\maketitle


\abstract{In this chapter, we discuss several aspects of the theory of strong interactions
in presence of a strong magnetic background. In particular, we summarize our results on 
the effect of the magnetic background on chiral symmetry restoration and deconfinement 
at finite temperature. Moreover, we compute the magnetic susceptibility of the chiral
condensate and the quark polarization at zero temperature. Our theoretical framework is 
given by chiral models: the Nambu-Jona-Lasinio (NJL), the Polyakov improved NJL (or PNJL)
and the Quark-Meson (QM) models. We also compare our results with the ones obtained
by other groups.}

\section{Introduction}
\label{sec:Introduction}
Quantum Chromodynamics (QCD) is the gauge theory of strong interactions. 
The understanding of its vacuum, and how it is modified by a large temperature
and/or a baryon density, is one of
the most intriguing aspects of modern physics. However, it is very
hard to get a full understanding of its properties, because its
most important characteristics, namely chiral symmetry breaking
and color confinement, have a non-perturbative origin, and the use
of perturbative methods is useless. One of the best strategies to
overcome this problem is offered by Lattice QCD simulations at
zero chemical potential
(see~\cite{deForcrand:2006pv,Aoki:2006br,Bazavov:2009zn,Cheng:2009be}
for several examples and see also references therein).  At
vanishing quark chemical potential, 
two crossovers take place in a broad range of temperatures; one for
quark deconfinement, and another one for the (approximate)
restoration of chiral symmetry.   

An alternative approach to the physics of strong interactions,
which is capable to capture some of the non-perturbative
properties of the QCD vacuum, is the use of models. 
Among them, we will consider here the Nambu-Jona Lasinio (NJL)
model~\cite{Nambu:1961tp}, see Refs.~\cite{revNJL} for
reviews. In this gluon-less model, which was inspired by the microscopic
theory of superconductivity, the QCD interactions are
replaced by effective interactions, which are built
in order to respect the global symmetries of QCD. Since 
gluons are absent in the NJL model, it is not a gauge theory.
However, it shares the global symmetries of the QCD action;
moreover, the parameters of the NJL model are fixed to reproduce
some phenomenological quantity of the QCD vacuum. Therefore, it is
the main characteristics of its phase diagram
should represent, at least qualitatively, those of QCD.

The other side of the NJL model is that it lacks confinement. 
The latter, in the case of a pure gauge theory, can be described in terms of the
center symmetry of the color gauge group and of the Polyakov
loop~\cite{Polyakovetal}, which is an order parameter for the
center symmetry. Motivated by this property, the Polyakov extended
Nambu-Jona Lasinio model (PNJL model) has been
introduced~\cite{Meisinger:1995ih,Fukushima:2003fw}, in which the
concept of statistical confinement replaces that of the true
confinement of QCD, and an effective interaction among the chiral
condensate and the Polyakov loop is achieved by a covariant
coupling of quarks with a background temporal gluon field. In the
literature there are several studies on the
PNJL model, 
see Refs.~\cite{Ratti:2005jh,Roessner:2006xn,Megias:2006bn,Sasaki:2006ww,
Ghosh:2007wy,Fukushima:2008wg,Fukushima:2008wg,Abuki:2008nm,Kahara:2010wh,
Sakai:2008py,Sakai:2009dv,Kashiwa:2009ki,Abuki:2008tx,Sasaki:2010jz,Hell:2008cc,
Kashiwa:2007hw} and references therein.

In this Chapter, we make use of the PNJL model to study the interplay
between chiral symmetry breaking and deconfinement in a {\em strong magnetic
background}. Moreover, we compute several quantities which are relevant
for the phenomenology of strong interactions physics in presence of a 
magnetic background. These topics are widely studied in the literature using many theoretical
approaches. Lattice studies on the
response to external magnetic (and chromo-magnetic) fields can be found
in~\cite{D'Elia:2010nq,Buividovich:2009my,Buividovich:2008wf,Cea:2002wx,Cea:2007yv,Bali:2012zg,Bali:2011qj}. 
Previous studies of QCD in magnetic fields, and of QCD-like theories as well, can be found in
Refs.~\cite{Klevansky:1989vi,Gusynin:1995nb,Klimenko:1990rh,Agasian:2008tb,Preis:2010cq}.
Self-consistent model calculations of
deconfinement pseudo-critical temperature in magnetic field have
been performed~\cite{Fukushima:2010fe,Mizher:2010zb,Campanelli:2009sc}. 

An important motivation for these kind of studies is phenomenological. 
In fact,
strong magnetic fields are produced in non-central heavy ion
collisions~\cite{Kharzeev:2007jp,Skokov:2009qp,Voronyuk:2011jd}. For example, 
at the energy scale for RHIC it is found $eB_{max} \approx 5 m_\pi^2$;
for collisions at the LHC energy scale $eB_{max} \approx 15 m_\pi^2$.
In this case, it has been argued that the non-trivial topological structure of
thermal QCD, namely the excitation of the strong sphalerons~\cite{Moore:2010jd}, locally changes the chirality
of quarks; this is reflected to event-by-event charge separation, 
a phenomenon which is dubbed Chiral Magnetic Effect (CME)
\cite{Kharzeev:2007jp,Buividovich:2009wi,Fukushima:2008xe}.
The possibility that the CME is observed in heavy ion collision experiments
has motivated the study of strong interactions in presence of a chirality imbalance
and a magnetic background,
see~\cite{Fukushima:2008xe,Fukushima:2010zza,Gatto:2011wc,Ruggieri:2011xc,Chernodub:2011fr,
Bayona:2011ab} and references therein. An experimental measurement of observables
connected to charge separation has been reported by the ALICE collaboration in~\cite{Abelev:2012pa}.
It is fair to say that realistic simulations
of heavy ion collisions show that the magnetic fields have a very short lifetime,
and decay before the local equilibrium is reached in the fireball.
Moreover, the magnetic field is highly inhomogeneous. Furthermore, electric fields
are produced beside the magnetic fields. Therefore, in order to describe
realistically hot matter produced in the collisions, one should take care of
the aforementioned details. However, for simplicity we neglect them,
and leave the (much harder) complete problem to future studies.

We mainly base the present Chapter on our previous works~\cite{Gatto:2010pt,Gatto:2010qs,Frasca:2011zn}.
We firstly discuss chiral symmetry restoration in a strong magnetic background
at finite temperature, using the PNJL model augmented with the eight-quark
interaction~\cite{Osipov:2006ev,Kashiwa:2006rc,Osipov:2005tq,Osipov:2007je}.
In this case we also compute the dressed Polyakov loop in a magnetic field.
The scenario which turns out from our calculations is compatible with that of the
magnetic catalysis, in which the magnetic field acts as a catalyzer
for chiral symmetry breaking. Moreover, we discuss on the role of the entanglement NJL vertex on the separation
between deconfinement and chiral symmetry restoration in the background field. 
Finally, we summarize our computation of the 
magnetic susceptibility of the chiral condensate and of quark polarization in 
a strong magnetic background at zero temperature. We base the latter analysis on the Quark-Meson (QM) model,
which offers the simplest renormalizable extension of the NJL model. 
Throughout this Chapter we consider QCD in the vacuum, that is at zero baryon
(as well as isospin) chemical potential. Computations at finite chemical potential
are present in the literature, as we will mention in the main body of the Chapter.


\section{The PNJL model with a magnetic background}
In this Section, we mainly summarize the results obtained in~\cite{Gatto:2010pt,Gatto:2010qs}.
We consider quark matter modeled by the following Lagrangian
density:
\begin{eqnarray}
{\cal L} &=& \bar q\left(i\gamma^\mu D_\mu - m_0\right)q
        + g_\sigma\left[(\bar q q)^2 + (\bar q i \gamma_5 \bm\tau q)^2\right] \nonumber \\
        &&+g_8\left[(\bar q q)^2 + (\bar q i \gamma_5 \bm\tau
        q)^2\right]^2~,
\label{eq:lagr}
\end{eqnarray}
which corresponds to the NJL lagrangian with multi-quark
interactions~\cite{Osipov:2006ev}. The covariant derivative embeds
the quark coupling to the external magnetic field and to the
background gluon field as well, as we will see explicitly below.
In Eq.~\eqref{eq:lagr}, $q$ represents a quark field in the
fundamental representation of color and flavor (indices are
suppressed for notational simplicity); $m_0$ is the bare quark
mass, which is fixed to reproduce the pion mass in the vacuum,
$m_\pi = 139$ MeV. Our interaction in Eq.~\eqref{eq:lagr} consists
of a four-quark term, whose coupling $g_\sigma$ has inverse mass
dimension of two, and an eight-quark term, whose coupling constant
$g_8$ has inverse mass dimension of eight.

We are considering the effect of a strong magnetic background on 
chiral symmetry restoration as well as deconfinement at finite temperature.
We assume the magnetic field to be along the positive $z-$axis; we chose
to work in the Landau gauge, specified by the vector potential $\bm A = (0,B x,0)$.

To compute a temperature for the deconfinement crossover, we use the
expectation value of the Polyakov loop, that we denote by $L$. In
order to compute $L$ we introduce a static, homogeneous and
Euclidean background temporal gluon field, $A_0 = iA_4 = i
\lambda_a A_4^a$, coupled minimally to the quarks via the QCD
covariant derivative~\cite{Fukushima:2003fw}. Then
\begin{equation}
L = \frac{1}{3}\text{Tr}_c\exp\left(i\beta\lambda_a A_4^a\right)~,
\end{equation}
where $\beta = 1/T$. In the Polyakov gauge, which is convenient
for this study, $A_0 = i\lambda_3 \phi + i \lambda_8 \phi^8$;
moreover, we work at zero quark chemical potential, therefore we
can take $L = L^\dagger$ from the beginning, which implies $A_4^8
= 0$. This choice is also motivated by the study
of~\cite{Mizher:2010zb}, where it is shown that the paramagnetic
contribution of the quarks to the thermodynamic potential induces
the breaking of the $Z_3$ symmetry, favoring the configurations
with a real-valued Polyakov loop.

Besides the Polyakov loop, it is interesting to compute the dressed Polyakov loop~\cite{Bilgici:2008qy}. 
In order to define this quantity, we work in a finite Euclidean volume with
temperature extension $\beta=1/T$. We take {\em twisted} fermion
boundary conditions along the compact temporal direction,
\begin{equation}
q(\bm x,\beta) = e^{-i\varphi}q(\bm x,0)~,~~~\varphi\in[0,2\pi]~,
\label{eq:phi}
\end{equation}
while for spatial directions the usual periodic boundary condition
is taken. The canonical antiperiodic boundary condition for the
quantization of fermions at finite temperature, is obtained by
taking $\varphi = \pi$ in the previous equation. The dual quark
condensate, $\tilde\Sigma_n$, is defined as
\begin{equation}
\tilde\Sigma_n(m,V) =
\int_0^{2\pi}\frac{d\varphi}{2\pi}\frac{e^{-i\varphi n}}{V}
\langle\bar q q\rangle_G~, \label{eq:Sn0}
\end{equation}
where $n$ is an integer. The expectation value
$\langle\bm\cdot\rangle_G$ denotes the path integral over gauge
field configurations. An important point is that in the
computation of the expectation value, the twisted boundary
conditions acts only on the fermion determinant; the gauge fields
are taken to be quantized with the canonical periodic boundary
condition.

Using a lattice regularization, it has been shown
in~\cite{Bilgici:2008qy} that Eq.~\eqref{eq:Sn0} can be expanded
in terms of loops which wind $n$ times along the compact time
direction. In particular, the case $n=1$ is called the {\em dressed
Polyakov loop}; it corresponds to a sum of loops winding just once
along the time direction. These correspond to the thin Polyakov
loop (the loop with shortest length) plus higher order loops, the
order being proportional to the length of the loop. Each higher
order loop is weighed by an inverse power of the quark mass.
Because of the weight, in the infinite quark mass limit only the
thin Polyakov loop survives; for this reason, the dressed Polyakov
loop can be viewed as a mathematical dressing of the thin loop, by
virtue of longer loops, the latter being more and more important
as the quark mass tends to smaller values.

If we denote by $z$ an element of the center of the color gauge
group, then $\tilde\Sigma_n \rightarrow
z^n\tilde\Sigma_n$. It then follows that, under the center of the
symmetry group $Z_3$, the dressed Polyakov loop $(n=1)$ is an order
parameter for the center symmetry, with the same transformation
rule of the thin Polyakov loop. Since the center symmetry is
spontaneously broken in the deconfinement phase and restored in
the confinement phase~\cite{Polyakovetal} (in presence of
dynamical quarks, it is only approximately restored), the dressed
Polyakov loop can be regarded as an order parameter for the
confinement-deconfinement transition as well.

\subsection{The one-loop quark propagator}
The evaluation of the bulk thermodynamic quantities requires we
compute the quantum effective action of the model. This cannot be
done exactly; hence we rely ourselves to the one-loop
approximation for the partition function, which amounts to take
the classical contribution plus the fermion determinant. 
At this level, the effect of the strong interactions is to 
modify the quark mass as follows:
\begin{equation}
M = m_0 - 2\sigma -4\sigma^3 g_8/g_\sigma^3~,
\label{eq:MASS}
\end{equation}
where $\sigma=g_\sigma\langle\bar q q\rangle$.
As a further simplification, we neglect condensation in the pseudoscalar channel.
We notice that the quark mass depends only on the sum of the $u$ and $d$ chiral condensates;
therefore the mean field quark mass does not depend on the flavor. 

To write the one-loop quark propagator in the background of the magnetic field
we make use of the Leung-Ritus-Wang
method~\cite{Ritus:1972ky}, which allows to expand the propagator
on the complete and orthonormal set made of the eigenfunctions of
a charged fermion in a homogeneous and static magnetic field. This
is a well known procedure, discussed many times in the literature,
see for
example~\cite{Elizalde:2000vz,Ferrer:2005vd,Fukushima:2009ft,Fukushima:2007fc};
therefore it is enough to quote the final result:
\begin{equation}
S_f(x,y) = \sum_{k=0}^\infty\int\frac{dp_0 dp_2 dp_3}{(2\pi)^4}
E_P(x)\Lambda_k \frac{1}{P\cdot\gamma - M}\bar{E}_P(y)~,
\label{eq:QP}
\end{equation}
where $E_P(x)$ corresponds to the eigenfunction of a charged
fermion in magnetic field, and $\bar{E}_P(x) \equiv
\gamma_0(E_P(x))^\dagger \gamma_0$. In the above equation,
\begin{equation}
P = (p_0 + i A_4,0,{\cal Q}\sqrt{2k|Q_f eB|},p_z)~,\label{eq:MB}
\end{equation}
where $k =0,1,2,\dots$ labels the $k^{\text{th}}$ Landau level,
and ${\cal Q} \equiv\text{sign}(Q_f)$, with $Q_f$ denoting the
charge of the flavor $f$; $\Lambda_k$ is a projector in Dirac
space which keeps into account the degeneracy of the Landau
levels; it is given by
\begin{equation}
\Lambda_k = \delta_{k0}\left[{\cal P_+}\delta_{{\cal Q},+1} +
{\cal P_-}\delta_{{\cal Q},-1}\right] + (1-\delta_{k0})I~,
\end{equation}
where ${\cal P}_{\pm}$ are spin projectors and $I$ is the identity
matrix in Dirac spinor indices. The propagator in
Eq.~\eqref{eq:QP} has a non-trivial color structure, due to the
coupling to the background gauge field, see Eq.~\eqref{eq:MB}.

It is useful to write down explicitely the expression of the chiral condensates
for the flavor $f$ with $f=u,d$. The chiral condensate is easily computed by taking
minus the trace of the $f$-quark propagator. It is easy to show that
the following equation holds:
\begin{eqnarray}
\langle\bar f f\rangle &=& - N_c\frac{|Q_f
eB|}{2\pi}\sum_{k=0}^\infty \beta_k \int\frac{dp_z}{2\pi}
\frac{M_f}{\omega_f} {\cal C}(L,\bar L, T|p_z,
k)~.\label{eq:CC}
\end{eqnarray}
Here,
\begin{eqnarray}
{\cal C}(L,\bar L, T|p_z, k) &=& U_\Lambda - 2{\cal N}(L,\bar L,
T|p_z, k)~,\label{eq:CCc}
\end{eqnarray}
and ${\cal N}$ denotes the statistically confining Fermi
distribution function,
\begin{eqnarray}
{\cal C}(L,\bar L, T|p_z, k) &=& \frac{1 + 2L e^{\beta\omega_f} +
Le^{2\beta\omega_f} }
  {1+3L e^{\beta\omega_f} + 3L e^{2\beta\omega_f} +
  e^{3\beta\omega_f}}~,\nonumber\\
  &&
\end{eqnarray}
where
\begin{equation}
\omega_f^2 = p_z^2 + 2|Q_f e B|k + M^2~.
\end{equation}
The first and the second addenda in the r.h.s. of
Eq.~\eqref{eq:CC} correspond to the vacuum and the thermal
fluctuations contribution to the chiral condensate, respectively.
The coefficient $\beta_k = 2 -\delta_{k0}$ keeps into account the
degeneracy of the Landau levels. The vacuum contribution is
ultraviolet divergent. In order to regularize it, we adopt a
smooth regulator $U_\Lambda$, which is more suitable, from the
numerical point of view, in our model calculation with respect to
the hard-cutoff which is used in analogous calculations without
magnetic field. We chose
\begin{equation}
U_\Lambda = \frac{\Lambda^{2N}}{\Lambda^{2N} + (p_z^2 + 2|Q_f e
B|k)^N}~. \label{eq:ULAMBDA}
\end{equation}


\subsection{The one-loop thermodynamic potential}
The one-loop thermodynamic potential of quark matter in external
fields has been discussed
in~\cite{Fukushima:2010fe,Campanelli:2009sc}, in the case of
canonical antiperiodic boundary conditions;
following~\cite{Kashiwa:2009ki}, it is easy to generalize it to
the more general case of twisted boundary conditions:
\begin{eqnarray}
\Omega &=& {\cal U}(L,\bar L, T) + \frac{\sigma^2}{g_\sigma} +
\frac{3\sigma^4 g_8}{g_\sigma^4}
        - \sum_{f=u,d}\frac{|Q_f e B|}{2\pi}
           \sum_{k}\beta_{k}\int_{-\infty}^{+\infty}\frac{dp_z}{2\pi}g_\Lambda(p_z,k)\omega_{k}(p_z)
  \nonumber \\
      &-&T\sum_{f=u,d}\frac{|Q_f e B|}{2\pi}\sum_{k}\beta_{k}
        \int_{-\infty}^{+\infty}\frac{dp_z}{2\pi}
        \log\left(1+3L e^{-\beta\cal{E}_-} + 3\bar{L}e^{-2\beta\cal{E}_-}+e^{-3\beta\cal{E}_-} \right)
      \nonumber \\
     &-&T\sum_{f=u,d}\frac{|Q_f e B|}{2\pi}\sum_{k}\beta_{k}
        \int_{-\infty}^{+\infty}\frac{dp_z}{2\pi}
        \log\left(1+3\bar L e^{-\beta\cal{E}_+} + 3Le^{-2\beta\cal{E}_+}+e^{-3\beta\cal{E}_+} \right)~.\nonumber\\
        &&
\label{eq:Om1}
\end{eqnarray}
In the previous equation
the arguments of the thermal exponentials are
defined as
\begin{equation}
{\cal E}_\pm = \omega_f(p_z) \pm \frac{i(\varphi-\pi)}{\beta}~,
\end{equation}
with $\varphi$ defined in Eq.~\eqref{eq:phi}.

The potential term $\mathcal{U}[L,\bar L,T]$ in Eq.~\eqref{eq:Om1}
is built by hand in order to reproduce the pure gluonic lattice
data~\cite{Ratti:2005jh}.  Among several different potential
choices~\cite{Schaefer:2009ui} we adopt the following logarithmic
form~\cite{Fukushima:2003fw,Ratti:2005jh},
\begin{equation}
  \frac{\mathcal{U}[L,\bar L,T]}{T^4} = -\frac{a(T)}{2}
  \bar L L + b(T)\ln\bigl[ 1-6\bar LL + 4(\bar L^3 + L^3)
  -3(\bar LL)^2 \bigr] \;,
\label{eq:Poly}
\end{equation}
with three model parameters (one of four is constrained by the
Stefan-Boltzmann limit),
\begin{equation}
 \begin{split}
 a(T) &= a_0 + a_1 \left(\frac{T_0}{T}\right)
 + a_2 \left(\frac{T_0}{T}\right)^2 , \\
 b(T) &= b_3\left(\frac{T_0}{T}\right)^3 \;.
 \end{split}
\label{eq:lp}
\end{equation}
The standard choice of the parameters reads~\cite{Ratti:2005jh};
\begin{equation}
 a_0 = 3.51\,, \quad a_1 = -2.47\,, \quad
 a_2 = 15.2\,, \quad b_3 = -1.75\,.
\end{equation}
The parameter $T_0$ in Eq.~\eqref{eq:Poly} sets the deconfinement
scale in the pure gauge theory, i.e.\ $T_c = 270$ \text{MeV}.

\section{Numerical results}
In this Section, we show our results. The main goal to achieve
numerically is the solution of the gap equations,
\begin{equation}
\frac{\partial\Omega}{\partial\sigma}=0~,~~~
\frac{\partial\Omega}{\partial L}=0~. \label{eq:GE}
\end{equation}
This is done by using a globally convergent algorithm with
backtrack~\cite{NumericalRecipes}. From the very definition of the
dressed Polyakov loop, Eq.~\eqref{eq:Sn0}, the twisted boundary
condition, Eq.~\eqref{eq:phi}, must be imposed only in
$D_\varphi$. Therefore, we firstly compute the expectation value
of the Polyakov loop and to the chiral condensate, taking
$\varphi=\pi$. Then, in order to compute the dressed Polyakov loop, we compute the $\varphi$-dependent
chiral condensate using the first of Eq.~\eqref{eq:GE}, keeping
the expectation value of the Polyakov loop fixed at its value at
$\varphi=\pi$~\cite{Kashiwa:2009ki}.

\begin{table}[t!]
\caption{\label{Tab:para}Parameters of the model for the two
choices of the UV-regulator.}
\begin{tabular}{ccccc}
 &$\Lambda$ (MeV)& $m_0$ (MeV) & $g_\sigma$ (MeV)$^{-2}$ & $g_8$ (MeV)$^{-8}$\\
\hline

$N=5$ &$588.657$ & 5.61& $5\times 10^{-6}$ & $6\times 10^{-22}$\\
\hline

\end{tabular}
\end{table}

Following~\cite{Gatto:2010pt,Gatto:2010qs} we report results obtained using the UV-regulator
with $N=5$. As expected,
in the other cases no different qualitative results are found; the parameter set is
specified in Table~\ref{Tab:para}. In the case $N=5$, they are
obtained by the requirements that the vacuum pion mass is $m_\pi =
139$ MeV, the pion decay constant $f_\pi = 92.4$ MeV and the
vacuum chiral condensate $\langle\bar u u\rangle
\approx(-241~\text{MeV})^3$. In this case, the chiral and
deconfinement pseudo-critical temperatures at zero magnetic field
are $T_0^\chi = T_0^P = 175$~\text{MeV}. 

The main effect of the eight-quark interaction in
Eq.~\eqref{eq:lagr} is to lower the pseudo-critical temperature of
the crossovers. This has been already discussed several times in
the literature~\cite{Kashiwa:2006rc,Osipov:2006ev}, in the context
of both the NJL and the PNJL models. Therefore, it is not
necessary to discuss it further here, while at the same time we
prefer to stress the results that have not been discussed yet.

In order to identify the pseudo-critical temperatures, we have
define the {\em effective susceptibilities} as
\begin{equation}
\chi_A = (m_\pi)^g\left|\frac{d A}{d
T}\right|~,~~~A=\sigma,P,\Sigma_1~. \label{eq:efs}
\end{equation}
Strictly speaking, the quantities defined in the previous equation
are not true susceptibilities. Nevertheless, they allow to
represent faithfully the pseudo-critical region, that is, the
range in temperature in which the various crossovers take place.
Therefore, for our purposes it is enough to compute these
quantities. In Equation~\eqref{eq:efs}, the appropriate power of
$m_\pi$ is introduced just for a matter of convenience, in order
to have a dimensionless quantity; therefore, $g=0$ if
$A=\sigma,\Sigma_1$, and $g=1$ if $A=P$.

\subsection{Condensates and dressed Polyakov loop}

\begin{figure*}
\begin{center}
\includegraphics[width=9cm]{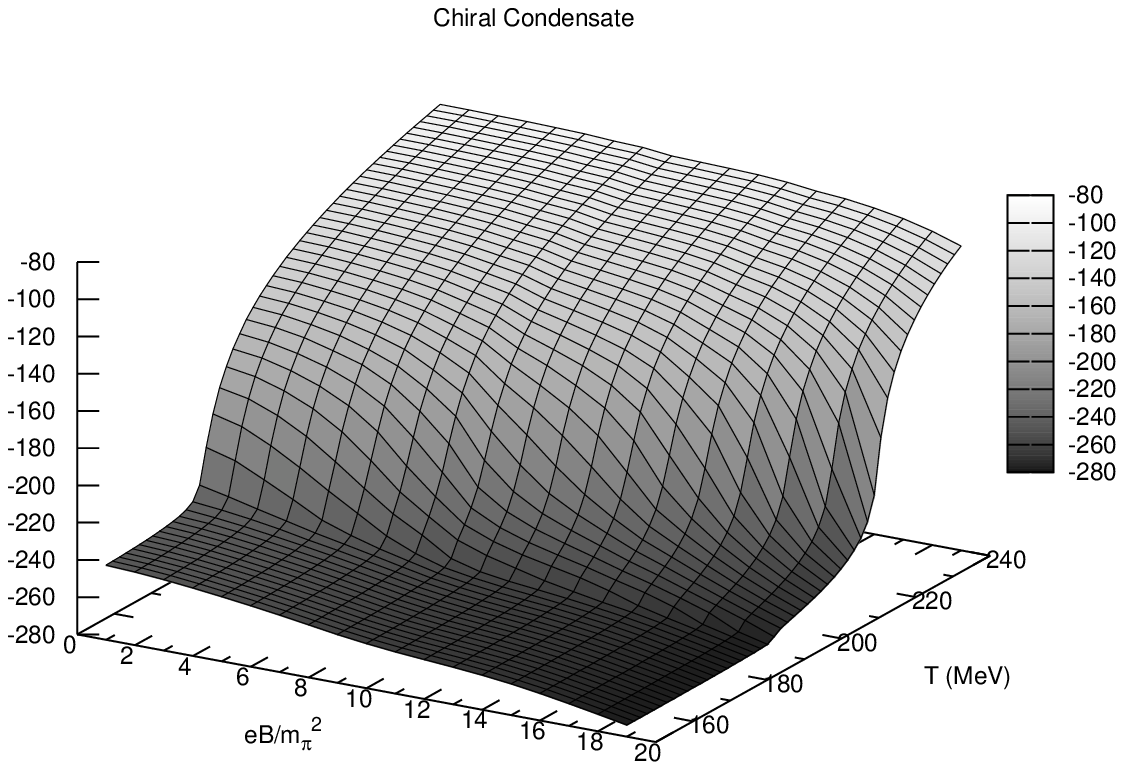}
\\
\includegraphics[width=9cm]{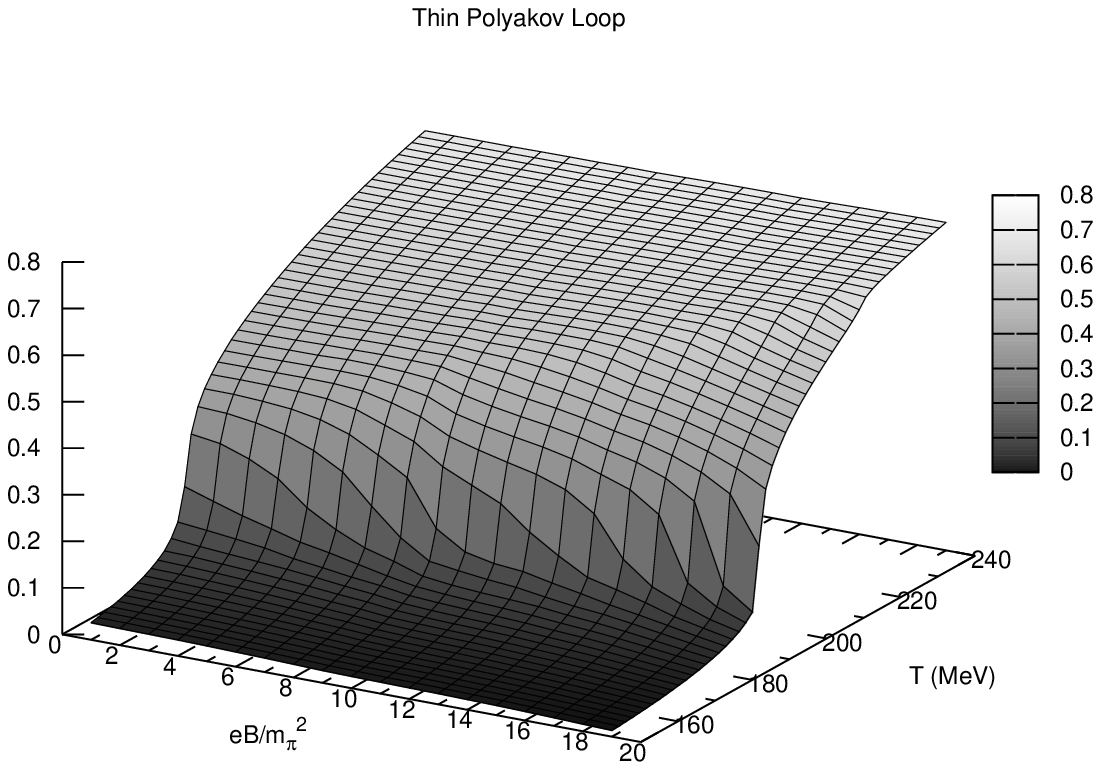}
\\
\includegraphics[width=9cm]{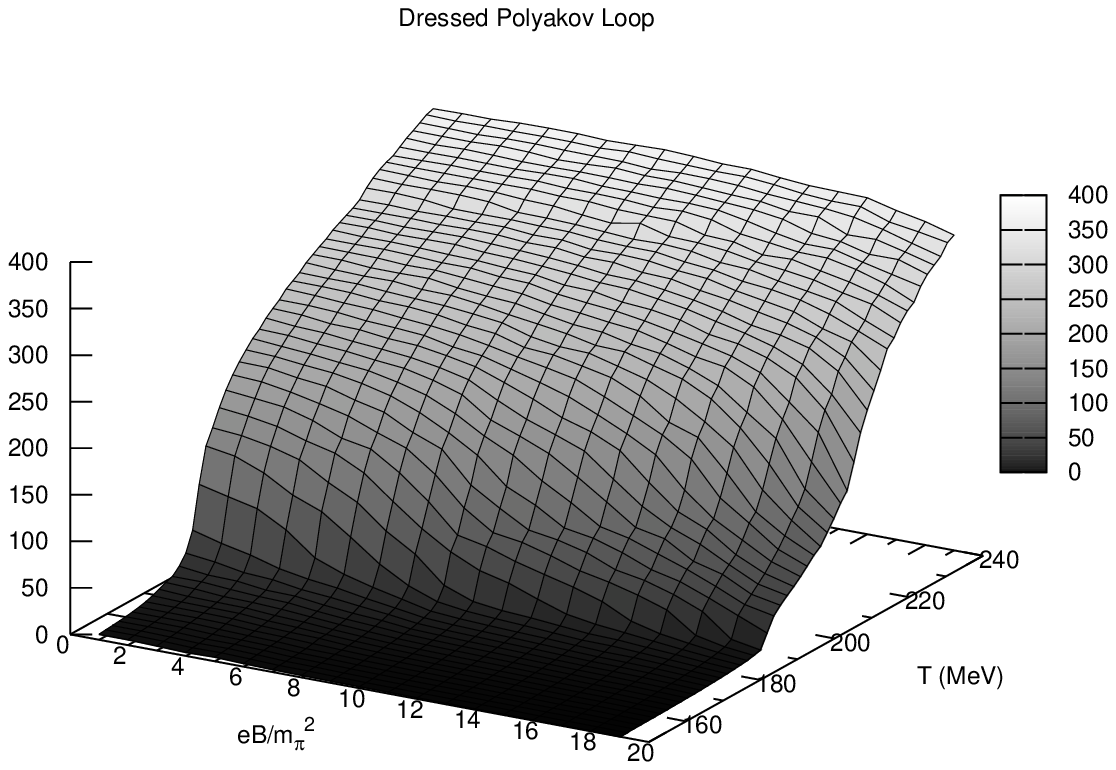}
\end{center}
\caption{\label{Fig:S3D} Chiral condensate,
Polyakov loop and dressed Polyakov loop as a function of
temperature and magnetic field, for the case $N=5$. From Ref.~\cite{Gatto:2010qs}.
Copyright(2012) by the American Physical Society.}
\end{figure*}

From now on, we fix $N=5$ unless specified. The results for this
case are collected in the form of surface plots in
Fig.~\ref{Fig:S3D}. In more detail, in the figure we plot the chiral condensate
$S\equiv(\sigma/2)^{1/3}$, the expectation value of the
Polyakov loop, and the dressed Polyakov loop $\Sigma_1$, as a
function of temperature and magnetic field.

\begin{figure*}
\begin{center}
\includegraphics[width=6cm]{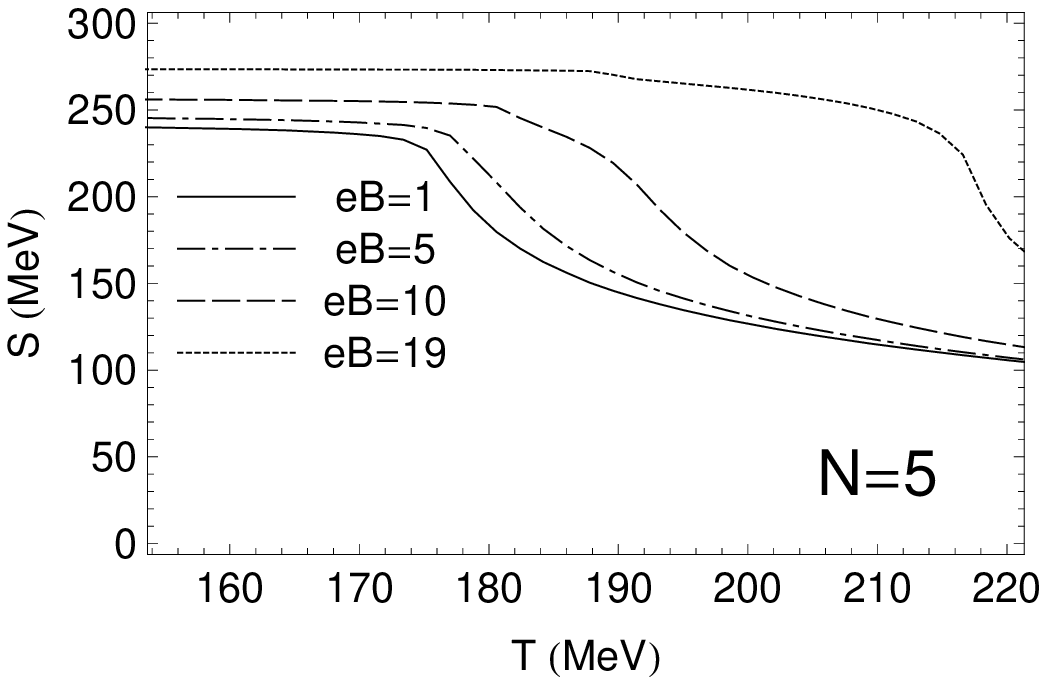}~~~\includegraphics[width=6cm]{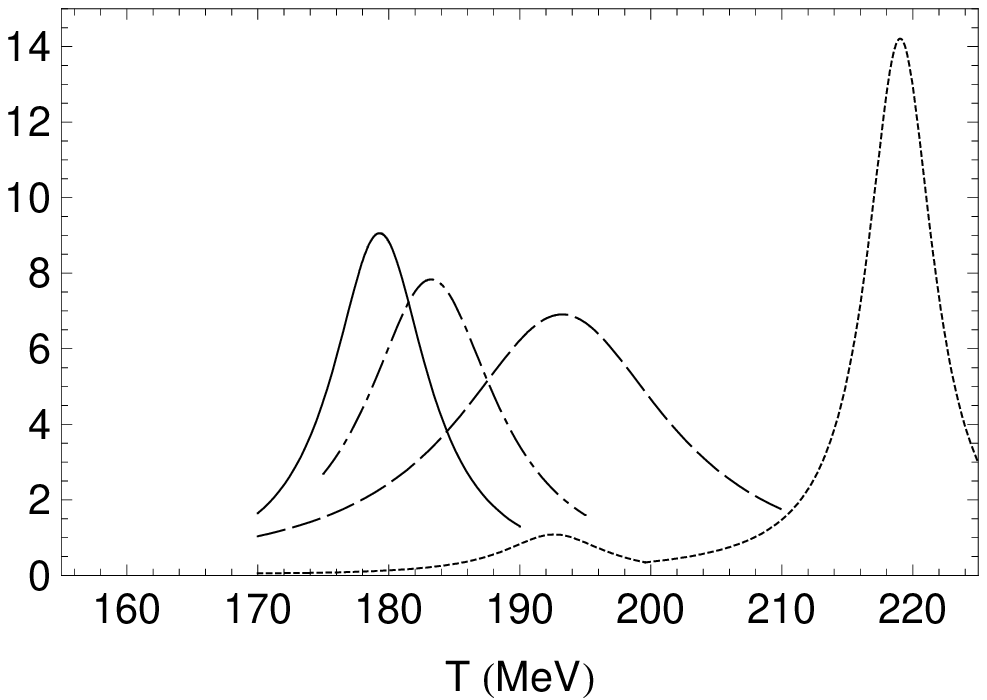}\\
\includegraphics[width=6cm]{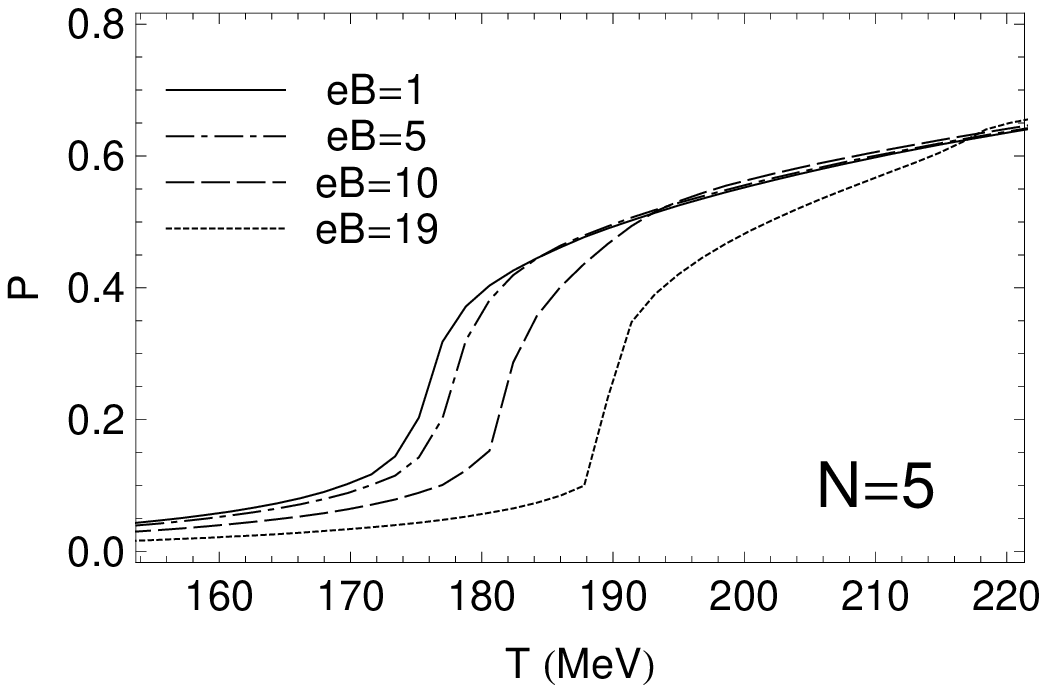}~~~\includegraphics[width=6cm]{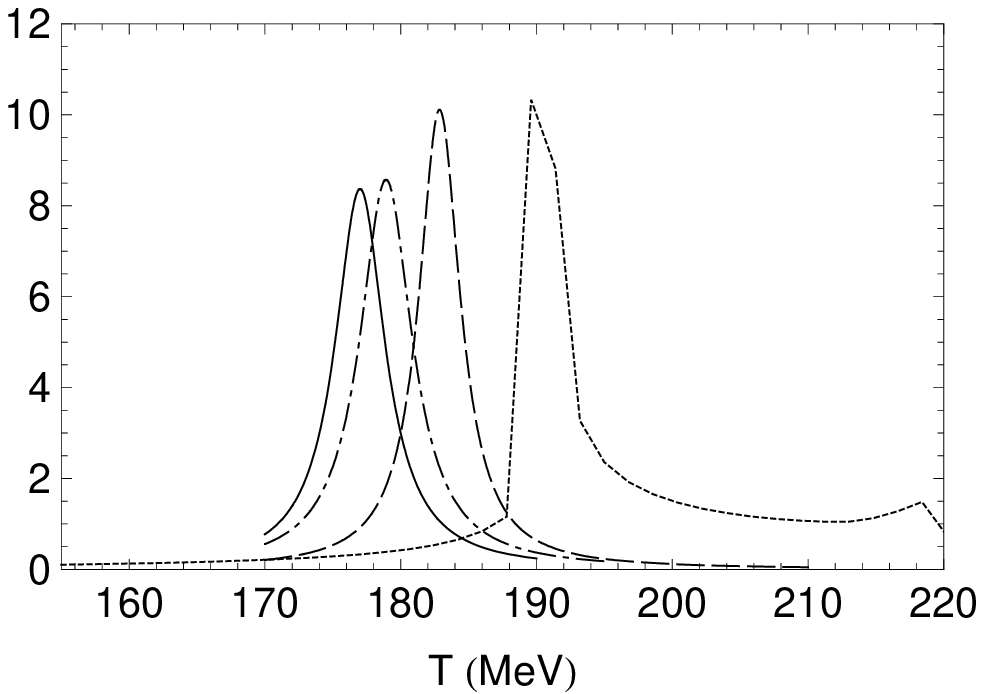}\\
\includegraphics[width=6cm]{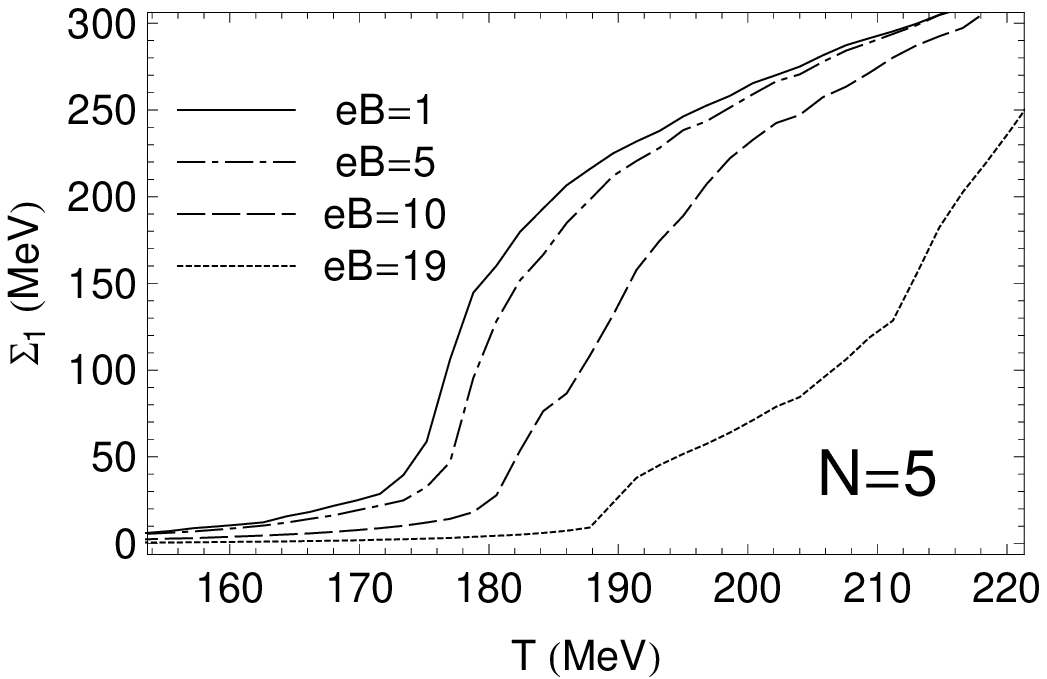}~~~\includegraphics[width=6cm]{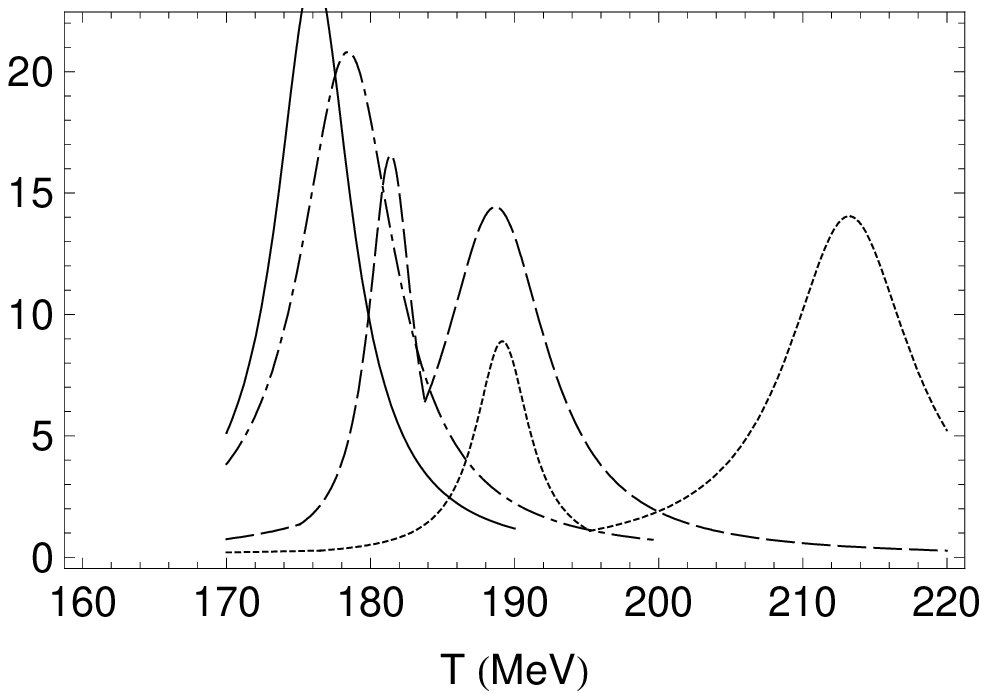}
\end{center}
\caption{\label{Fig:cN5}{\em Left panel}. Chiral condensate
$S$ (upper panel), Polyakov loop
(middle panel) and $\Sigma_1$ (lower panel) as a function of
temperature, for several values of the applied magnetic field
strength, measured in units of $m_\pi^2$. 
 {\em Right panel}. Effective
susceptibilities, defined in Eq.~\eqref{eq:efs}, as a function of
temperature, for several values of $eB$. Conventions for lines are
the same as in the left panel. From Ref.~\cite{Gatto:2010qs}.
Copyright(2012) by the American Physical Society.}
\end{figure*}

We slice the surface plots in Fig.~\ref{Fig:S3D} at
fixed value of the magnetic field strength, and show the results
in Fig.~\ref{Fig:cN5}, where we plot the chiral condensate
(upper panel), the Polyakov
loop (middle panel) and $\Sigma_1$ (lower panel) as a function of
temperature, for several values of the applied magnetic field
strength, measured in units of $m_\pi^2$. In the right panel, we
plot fits of the effective susceptibilities in the critical
regions, as a function of temperature. The fits are obtained from
the raw data, using Breit-Wigner-like fitting functions. 

The qualitative behavior of the chiral condensate, and of the
Polyakov loop as well, is similar to that found in a previous
study within the PNJL model in the chiral
limit~\cite{Fukushima:2010fe}. Quantitatively, the main difference
with the case of the chiral limit, is that in the latter the
chiral restoration at large temperature is a true second order
phase transition (in other model calculations it has been reported
that the phase transition might become of the first order at very
large magnetic field strengths~\cite{Agasian:2008tb}). On the
other hand, in the case under investigation, chiral symmetry is
always broken explicitly because of the bare quark masses; as a
consequence, the second order phase transition is replaced by a
crossover.

Another aspect is that the Polyakov loop crossover
temperature is less sensitive to the strength of the magnetic
field than the same quantity computed for the chiral condensate.
It is useful, for illustration purpose, to quantify the net shift
of the pseudo-critical temperatures, for the largest value of
magnetic field we have studied, $eB = 19 m_\pi^2$. In this case,
if we take $N=5$, then the two
crossovers occur simultaneously at $eB=0$, at the temperature
$T_0^\chi = T_0^P = 175$~\text{MeV}; for $eB = 19 m_\pi^2$, we
find $T_\chi = 219$~\text{MeV} and $T_P = 190$~\text{MeV}.
Therefore, the chiral crossover is shifted approximately by
$25.1\%$, to be compared with the more modest shift of the
Polyakov loop crossover, which is $\approx 8.6\%$. This aspect
will be discussed further in the next Section, in which we will
comment on the possibility of entanglement between the NJL coupling
at finite temperature and the Polyakov loop.

In the lower panels of Figures.~\ref{Fig:S3D} and~\ref{Fig:cN5},
we plot the dressed Polyakov loop as a function of temperature,
for several values of $eB$. We have normalized $\Sigma_1$ 
multiplying the one defined in~\cite{Bilgici:2008qy} by the NJL coupling constant.
For small values of $eB/m_\pi^2$, the behavior of $\Sigma_1$ as
temperature is increased, is qualitatively similar to that at
$eB=0$, which has been discussed within effective models
in~\cite{Kashiwa:2009ki,Mukherjee:2010cp}. In particular, the
dressed Polyakov loop is very small for temperatures below the
pseudo-critical temperature of the simultaneous crossover. Then,
it experiences a crossover in correspondence of the simultaneous
Polyakov loop and chiral condensate crossovers. It eventually
saturates at very large temperature (for example,
in~\cite{Kashiwa:2009ki} the saturation occurs at a temperature of
the order of $0.4$ GeV, in agreement with the results
of~\cite{Mukherjee:2010cp}). However, we do not push up our
numerical calculation to such high temperature, because we expect
that the effective model in that case is well beyond its range of
validity.

As we increase the value of $eB$, as noticed previously, we
observe a tiny splitting of the chiral and the Polyakov loop
crossovers. Correspondingly, the qualitative behavior of the
dressed Polyakov loop changes dramatically: the range of
temperature in which the $\Sigma_1$ crossover takes place is
enlarged, if compared to the thin temperature interval in which
the crossover takes place at the lowest value of $eB$ (compare the
solid and the dotted lines in Fig.~\ref{Fig:cN5}, as well as the
the lower panel of Fig.~\ref{Fig:S3D}).

On passing, we discuss briefly the effective susceptibility, $d\Sigma_1/dT$, plotted in the lower
right panel of Fig.~\ref{Fig:cN5}, since its qualitative behavior is very
interesting. We observe a double peak structure, which we
interpret as the fact that the dressed Polyakov loop is capable to
feel (and hence, describe) both the crossovers. If we were to
interpret $\Sigma_1$ as the order parameter for deconfinement, and
the temperature with the largest susceptibility as the crossover
pseudo-critical temperature, then we obtain almost simultaneous
crossover even for very large magnetic field. 


\subsection{Entanglement of NJL coupling and Polyakov loop}
In~\cite{Kondo:2010ts} it has been
shown that the NJL vertex can be deduced under some assumption from the QCD action;
following this derivation a non-local structure of the interaction turns out. 
An analogous conclusion is achieved
in~\cite{Frasca:2008zp}. More important for our study, the NJL
vertex acquires a non-trivial dependence on the phase of the
Polyakov loop. Therefore, in the model we consider here, it is
important to keep into account this dependence. Here we follow 
the phenomenological ansatz introduced in~\cite{Sakai:2010rp}, that
is
\begin{equation}
G = g_\sigma\left[1 - \alpha_1 L\bar{L} -\alpha_2(L^3 +
\bar{L}^3)\right]~,\label{eq:Run}
\end{equation}
and we take $L=\bar{L}$. Moreover, we mainly discuss here
the case without 8-quark interaction.
The model with coupling constant specified in Eq.~\eqref{eq:Run}
is named Entangled-Polyakov improved-NJL model (EPNJL in the following)~\cite{Sakai:2010rp},
since the vertex describes an entanglement between Polyakov loop 
and the interaction responsible for chiral symmetry breaking.

The functional form in
the above equation is constrained by $C$ and extended $Z_3$
symmetry. We refer to~\cite{Sakai:2010rp} for a more detailed
discussion. The numerical values of $\alpha_1$ and $\alpha_2$ have
been fixed in~\cite{Sakai:2010rp} by a best fit of the available
Lattice data at zero and imaginary chemical potential of
Ref.~\cite{D'Elia:2009qz}, which have been confirmed recently
in~\cite{Bonati:2010gi}. In particular, the fitted data are the
critical temperature at zero chemical potential, and the
dependence of the Roberge-Weiss endpoint on the bare quark mass.

The values $\alpha_1 = \alpha_2 \equiv \alpha = 0.2 \pm 0.05$ have
been obtained in~\cite{Sakai:2010rp} using a hard cutoff
regularization scheme. We will focus mainly on the case
$\alpha=0.2$ as in~\cite{Sakai:2010rp}. In~\cite{Gatto:2010pt} we have verified that in
the regularization scheme with the smooth cutoff, the results are in quantitative
agreement with those of~\cite{Sakai:2010rp}. There, a detailed discussion of the role
of $\alpha$ can be found as well (we will skip this discussion in this Chapter).

\begin{figure}[t!]
\begin{center}
\includegraphics[width=7cm]{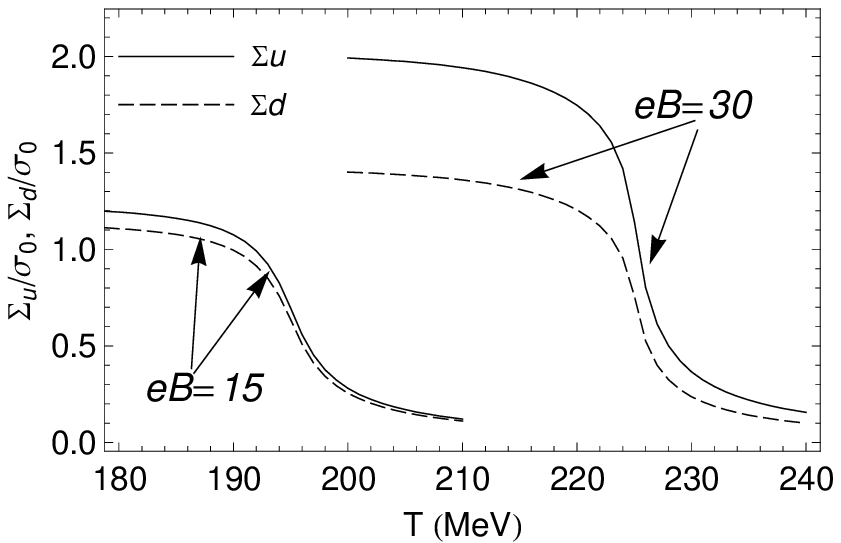}\\
\includegraphics[width=7cm]{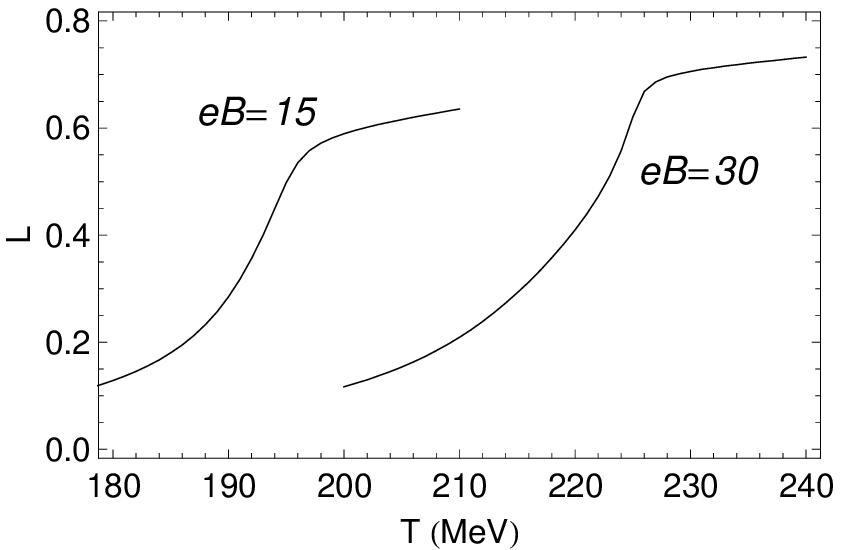}
\end{center}
\caption{\label{Fig:UDCC} {\em Upper panel:} Chiral condensates of
$u$ and $d$ quarks as functions of temperatures in the
pseudo-critical region, at $eB = 15 m_\pi^2$ and $eB = 30
m_\pi^2$. Condensates are measured in units of their value at zero
magnetic field and zero temperature, namely $\sigma_0 =
(-253~\text{MeV})^3$. {\em Lower panel:} Polyakov loop expectation
value as a function of temperature, at $eB = 15 m_\pi^2$ and $eB =
30 m_\pi^2$. Data correspond to $\alpha=0.2$. From Ref.~\cite{Gatto:2010qs}. 
Copyright(2012) by American Physical Society.}
\end{figure}

We plot in Fig~\ref{Fig:UDCC} the chiral
condensates of $u$ and $d$ quarks as a function of temperature, at
$eB = 15 m_\pi^2$ and $eB = 30 m_\pi^2$. In the lower panel of the figure,
we plot the expectation value of the Polyakov loop as a function of temperature.
The condensates are
measured in units of their value at zero magnetic field and zero
temperature, namely $\sigma_0 \equiv \langle\bar uu\rangle =
\langle\bar dd\rangle = (-253~\text{MeV})^3$. They are computed by
a two-step procedure: firstly we find the values of $\sigma$ and
$L$ that minimize the thermodynamic potential; then, we make use
of Eq.~\eqref{eq:CC} to compute the expectation values of $\bar
uu$ and $\bar dd$ in magnetic field. If we measure the strength of
the crossover by the value of the peak of $|d\sigma/dT|$, it is
obvious from the Figure that the chiral crossover becomes stronger
and stronger as the strength of the magnetic field is increased,
in agreement with~\cite{D'Elia:2010nq}.

The results in Fig~\ref{Fig:UDCC} show that, identifying the deconfinement
crossover with the temperature $T_L$ at which $dL/dT$ is maximum,
and the chiral crossover with the temperature $T_\chi$ at which
$|d\sigma/dT|$ is maximum, the two temperatures
are very close also in a strong magnetic field. 
From the model point of view, it is
easy to understand why deconfinement and chiral symmetry
restoration are entangled also in strong magnetic field. As a
matter of fact, using the data shown in Fig~\ref{Fig:UDCC}, it is
possible to compute the NJL coupling constant in the
pseudo-critical region, which turns out to decrease of the $15\%$ as a
consequence of the deconfinement crossover. Therefore, the
strength of the interaction responsible for the spontaneous chiral
symmetry breaking is strongly affected by the deconfinement, with
the obvious consequence that the numerical value of the chiral
condensate drops down and the chiral crossover takes place. We
have verified that the picture remains qualitatively and
quantitatively unchanged if we perform a calculation at $eB=30
m_\pi^2$. In this case, we find $T_L = 224$ MeV and $T_\chi = 225$
MeV.

This result can be compared with the previous
calculations~\cite{Fukushima:2010fe}, described also in the previous Section, in which 
the Polyakov loop dependence of the NJL coupling constant was not included.
In~\cite{Fukushima:2010fe} we worked in the chiral limit and we
observed that the Polyakov loop crossover in the PNJL model is
almost insensitive to the magnetic field; on the other hand, the
chiral phase transition temperature was found to be very sensitive
to the strength of the applied magnetic field, in agreement with
the well known magnetic catalysis
scenario~\cite{Klevansky:1989vi}. This model prediction has been
confirmed within the Polyakov extended quark-meson model
in~\cite{Mizher:2010zb}, when the contribution from the vacuum
fermion fluctuations to the energy density is kept into
account~\footnote{If the vacuum corrections are neglected, the
deconfinement and chiral crossovers are found to be coincident
even in very strong magnetic fields~\cite{Mizher:2010zb}, but the
critical temperature decreases as a function of $eB$; this
scenario is very interesting theoretically, but it seems
excluded from the recent Lattice
simulations~\cite{D'Elia:2010nq}.}; we then obtained a similar
result in~\cite{Gatto:2010qs}, in which we turned from the chiral
to the physical limit at which $m_\pi = 139$ MeV, and introduced
the 8-quark term as well (PNJL$_8$ model, according to the
nomenclature of~\cite{Sakai:2010rp}). The comparison with the
results of the PNJL$_8$ model of~\cite{Gatto:2010qs} is
interesting because the model considered there, was tuned in order
to reproduce the Lattice data at zero and imaginary chemical
potential~\cite{Sakai:2009dv}, like the model we use in this
study. Therefore, they share the property of describing the QCD
thermodynamics at zero and imaginary chemical potential; it is
therefore instructive to compare their predictions at finite $eB$.

For concreteness, in~\cite{Gatto:2010qs} we found $T_P =185$ MeV
and $T_\chi = 208$ MeV at $eB=19 m_\pi^2$, corresponding to a
split of $\approx 12\%$. On the other hand, in the present
calculation we measure a split of $\approx 1.5\%$ at the largest
value of $eB$ considered. Therefore, the results of the two models
are in slight quantitative disagreement; this disagreement is then
reflected in a slightly different phase diagram. We will draw the
phase diagram of the two models in a next Section; however, since
now it is easy to understand what the main difference consists in:
the PNJL$_8$ model predicts some window in the $eB-T$ plane in
which chiral symmetry is still broken by a chiral condensate, but
deconfinement already took place. In the case of the EPNJL model,
this window is shrunk to a very small one, because of the
entanglement of the two crossovers at finite $eB$. On the other
hand, it is worth to stress that the two models share an important
qualitative feature: both chiral restoration and deconfinement
temperatures are enhanced by a strong magnetic field; the latter conclusion is in
qualitative agreement with the Lattice
data of D'Elia {\em et al.}~\cite{D'Elia:2010nq}, but in disagreement
with the more recent data of the Wuppertal-Budapest group~\cite{Bali:2012zg,Bali:2011qj}.
We will come back to a comparison with Lattice data, as well as with other 
computations, in the next Section.

\section{Phase diagram in the $eB-T$ plane}

\begin{figure}[t!]
\begin{center}
\includegraphics[width=7cm]{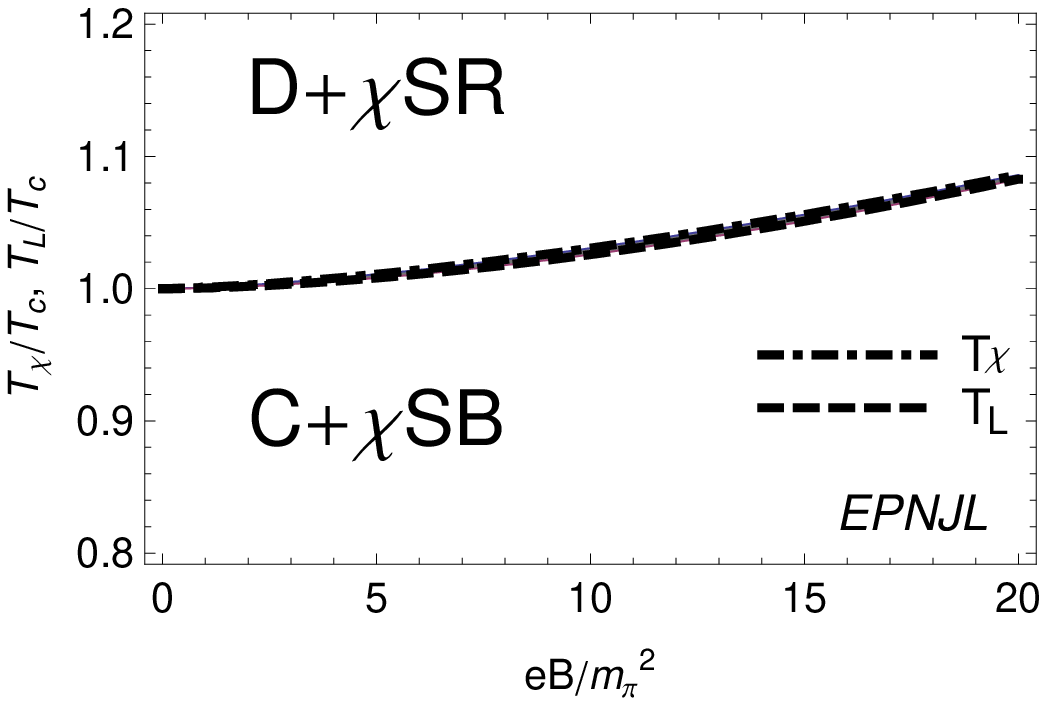}\\
\includegraphics[width=7cm]{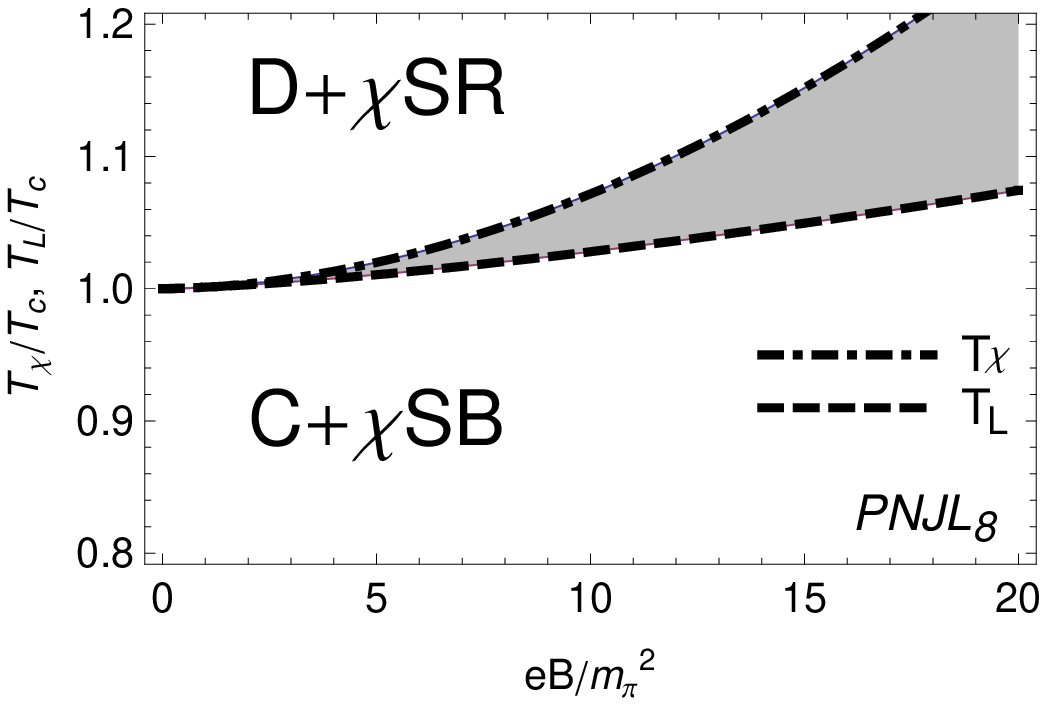}
\end{center}
\caption{\label{Fig:PD2} {\em Upper panel:} Phase diagram in the
$eB-T$ plane for the EPNJL model. Temperatures on the vertical
axis are measured in units of the pseudo-critical temperature for
deconfinement at $eB=0$, namely $T_c = 185.5$ MeV. {\em Lower
panel:} Phase diagram in the $eB-T$ plane for the PNJL$_8$ model.
Temperatures on the vertical axis are measured in units of the
pseudo-critical temperature for deconfinement at $eB=0$, namely
$T_c = 175$ MeV. In both the phase diagrams, $T_\chi$, $T_L$
correspond to the chiral and deconfinement pseudo-critical
temperatures, respectively. The grey shaded region denotes the
portion of phase diagram in which hot quark matter is deconfined
and chiral symmetry is still broken spontaneously. From Ref.~\cite{Gatto:2010qs}. 
Copyright(2012) by American Physical Society.}
\end{figure}

In Fig.~\ref{Fig:PD2} we collect our
data on the pseudo-critical temperatures for deconfinement and
chiral symmetry restoration, in the form of a phase diagram in the
$eB-T$ plane. In the upper panel we show the results obtained
within the EPNJL model; in the lower panel, we plot the results of
the PNJL$_8$ model, that are obtained using the fitting functions
computed in~\cite{Gatto:2010qs}. In the figure, the magnetic field
is measured in units of $m_\pi^2$; temperature is measured in
units of the deconfinement pseudo-critical temperature at zero
magnetic field, namely $T_{B=0} = 185.5$ MeV for the EPNJL model,
and $T_{B=0} = 175$ MeV for the PNJL$_8$ model. For any value of
$eB$, we identify the pseudo-critical temperature with the peak of
the effective susceptibility.

It should be kept in mind, however, that the definition of a
pseudo-critical temperature in this case is not unique, because of
the crossover nature of the phenomena that we describe. Other
satisfactory definitions include the temperature at which the
order parameter reaches one half of its asymptotic value (which
corresponds to the $T\rightarrow 0$ limit for the chiral
condensate, and to the $T\rightarrow +\infty$ for the Polyakov
loop), and the position of the peak in the true susceptibilities.
The expectation is that the critical temperatures computed in
these different ways differ from each other only of few percent.
This can be confirmed concretely using the data in
Fig.~\ref{Fig:UDCC} at $eB=30 m_\pi^2$. Using the peak of the
effective susceptibility we find $T_\chi = 225$ MeV and $T_L =
224$ MeV; on the other hand, using the half-value criterion, we
find $T_\chi = 227$ MeV and $T_L = 222$ MeV, in very good
agreement with the previous estimate. Therefore, the qualitative
picture that we derive within our simple calculational scheme,
namely the entanglement of the two crossovers in a strong magnetic
field, should not be affected by using different definitions of
the critical temperatures.

Firstly we focus on the phase diagram of the EPNJL model. In the upper panel of
Fig.~\ref{Fig:PD2}, the dashed and dot-dashed lines correspond to
the deconfinement and chiral symmetry restoration pseudo-critical
temperatures, respectively. As a
consequence of the entanglement, the two crossovers stay closed
also in very strong magnetic field, as we have already discussed
in the previous Section. The grey region in the Figure denotes a
phase in which quark matter is (statistically) deconfined, but
chiral symmetry is still broken. According
to~\cite{Cleymans:1986cq,Kouno:1988bi}, we can call this phase
Constituent Quark Phase (CQP).

On the lower panel of Fig.~\ref{Fig:PD2} we have drawn the phase
diagram for the PNJL$_8$ model based on Ref.~\cite{Gatto:2010qs}
and discussed in the previous Section.
The most astonishing feature of the phase diagram of
the PNJL$_8$ model is the entity of the split among the
deconfinement and the chiral restoration crossover. The difference
with the result of the EPNJL model is that in the former, the
entanglement with the Polyakov loop is neglected in the NJL
coupling constant. As we have already mentioned in the previous
Section, the maximum amount of split that we find within the EPNJL
model, at the largest value of magnetic field considered here, is
of the order of $2\%$; this number has to be compared with the
split at $eB=20 m_\pi^2$ in the PNJL$_8$ model, namely $\approx
12\%$. The larger split causes a considerable portion of the phase
diagram to be occupied by the CQP.


\subsection{Comparison with other computations}
In this Section we summarize the main results obtained in the literature,
comparing them with the scenario depicted in our works.
Before going ahead, it is useful to summarize the two main results obtained within
our one-loop computations:
\begin{itemize}
 \item The critical temperature for chiral symmetry restoration is {\em increased} by an external magnetic field;
 \item The split between deconfiment and chiral symmetry restoration temperatures in a strong magnetic background can 
 {\em be reduced} if the entanglement vertex is considered.
\end{itemize}
The first conclusion is in agreement with most of the computations: 
calculations based on the quark-meson-model with and without quantum 
fluctuations~\cite{Andersen:2012bq,Andersen:2011ip,Mizher:2010zb,Skokov:2011ib,Fukushima:2012xw}, 
on chiral perturbation theory at finite temperature~\cite{Andersen:2012zc},
on the PNJL model~\cite{Fukushima:2010fe,Kashiwa:2011js}, 
on the holographic correspondence~\cite{Bergman:2008sg,Johnson:2008vna,Zayakin:2008cy,Preis:2010cq}.

Besides models predictions, lattice computations of the critical temperatures in a magnetic background have been performed~\cite{D'Elia:2010nq,Bali:2012zg,Bali:2011qj}. 
The computations of~\cite{D'Elia:2010nq} have been performed with a pion mass of the order
of $400$ MeV, hence a little bit far from the physical limit. In this case, both the chiral symmetry
restoration and the deconfimenent temperatures are measured, and they
are found to increase slightly with the magnetic field strength; moreover, the two 
transitions seem to be entangled even at the largest value of the magnetic field considered
in the study.  On the other hand, in~\cite{Bali:2012zg,Bali:2011qj} quark masses are chosen
such that the lattice pion has its physical mass; in this case a non-trivial dependence of the critical temperature
for chiral symmetry restoration on the magnetic field strength and the quark mass is found.
In more detail, for the up and down quark condensates at the physical limit, 
the critical temperature decreases with the magnetic field strength.
On the other hand, the critical temperature for the strange quark condensate
is increased by the magnetic field, in agreement with the magnetic catalysis
scenario. A measurement of the deconfinement temperature has not been performed.
The results of~\cite{Bali:2012zg,Bali:2011qj} are quite surprising since they reveal 
an unexpected role of the quark mass on the curvature of the critical line in the $eB-T$
plane.

Computations within the MIT bag model~\cite{Fraga:2012fs} do not have direct access to the
chiral symmetry restoration, but to the deconfinement temperature. Within this model it is found
that the critical temperature for deconfinement is a decreasing function of the external
magnetic field strength. This conclusion is in agreement with a previous
computation~\cite{Agasian:2008tb}. Furthermore, a decreasing temperature
is found also within the quark-meson model if the fermion vacuum contribution is neglected~\cite{Mizher:2010zb}.
If deconfinement and chiral symmetry restoration are entangled
at finite magnetic field, then the MIT model based study would give some hint on the mechanism
which makes the temperature for chiral symmetry restoration in a magnetic background lower
than that at zero field. However, if this is the case, then the role of the quark mass
on the dependence of the critical temperature on the magnetic field should be transparent.
In our opinion, more study is needed to understand the puzzling behavior of $T_c$ as a function
of the magnetic field found on the Lattice: beside model computations, independent Lattice
simulations should be performed, in order to confirm the results of~\cite{Bali:2012zg,Bali:2011qj}.

\section{Polarization of the quark condensate}

It has been realized that external fields can induce QCD
condensates that are absent otherwise~\cite{Ioffe:1983ju}. Here
we focus on the magnetic moment,
$\langle\bar f \Sigma^{\mu\nu} f\rangle$ where $f$ denotes the
fermion field of the flavor $f-$th, and $\Sigma^{\mu\nu} =
-i(\gamma^\mu \gamma^\nu - \gamma^\nu\gamma^\mu)/2$. At small
fields one can write, according to~\cite{Ioffe:1983ju},
\begin{equation}
\langle\bar f \Sigma^{\mu\nu} f\rangle = \chi\langle\bar f
f\rangle Q_f |eB|~, \label{eq:chiDef}
\end{equation}
and $\chi$ is a constant independent of flavor, which is dubbed
magnetic susceptibility of the quark condensate.
In~\cite{Ioffe:1983ju} it is proved that the role of the
condensate~\eqref{eq:chiDef} to QCD sum rules in external fields
is significant, and it cannot be ignored. The quantity $\chi$ has
been computed by means of special sum
rules~\cite{Ioffe:1983ju,Belyaev:1984ic,Balitsky:1985aq,Ball:2002ps,Rohrwild:2007yt},
OPE combined with Pion Dominance~\cite{Vainshtein:2002nv},
holography~\cite{Gorsky:2009ma,Son:2010vc}, instanton vacuum
model~\cite{Kim:2004hd}, analytically from the zero mode of the
Dirac operator in the background of a $SU(2)$
instanton~\cite{Ioffe:2009yi}, and on the Lattice in two color
quenched simulations at zero and finite
temperature~\cite{Buividovich:2009ih}. It has also been suggested
that in the photoproduction of lepton pairs, the interference of
the Drell-Yan amplitude with the amplitude of a process where the
photon couples to quarks through its chiral-odd distribution
amplitude, which is normalized to the magnetic susceptibility of
the QCD vacuum, is possible~\cite{Pire:2009ap}. This interference
allows in principle to access the chiral odd transversity parton
distribution in the proton. Therefore, this quantity is
interesting both theoretically and phenomenologically. The several
estimates, that we briefly review in Section III, lead to the
numerical value of $\chi$ as follows:
\begin{equation}
\chi\langle\bar ff\rangle = 40-70~\text{MeV}~. \label{eq:i1}
\end{equation}

A second quantity, which embeds non-linear effects at large
fields, is the polarization, $\mu_f$, defined as
\begin{equation}
\mu_f = \left|\frac{\Sigma_f}{\langle\bar f f\rangle}\right|~,~~~
\Sigma_f = \langle\bar f\Sigma^{12}f\rangle~, \label{eq:muDef}
\end{equation}
which has been computed on the Lattice
in~\cite{Buividovich:2009ih} for a wide range of magnetic fields,
in the framework of two-color QCD with quenched fermions. At small
fields $\mu_f = |\chi Q_f eB|$ naturally; at large fields,
non-linear effects dominate and an interesting saturation of
$\mu_f$ to the asymptotic value $\mu_\infty = 1$ is measured.
According to~\cite{Buividovich:2009ih} the behavior of the
polarization as a function of $eB$ in the whole range examined,
can be described by a simple inverse tangent function. Besides,
magnetization of the QCD vacuum has been computed in the strong
field limit in~\cite{Cohen:2008bk} using perturbative QCD, where
it is found it grows as $B\log B$.

In~\cite{Frasca:2011zn} we compute the magnetic susceptibility of the
quark condensate by means of the NJL and the QM models. This study
is interesting because in the chiral models, it is possible to
compute self-consistently the numerical values of the condensates
as a function of $eB$, once the parameters are fixed to reproduce
some characteristic of the QCD vacuum. We firstly perform a
numerical study of the problem, which is then complemented by some
analytic estimate of the same quantity within the renormalized QM
model. Moreover, we compute the polarization of quarks at small as
well as large fields, both numerically and analytically. In
agreement with the Lattice results~\cite{Buividovich:2009ih}, we
also measure a saturation of $\mu_f$ to one at large fields, in
the case of the effective models. Our results push towards the
interpretation of the saturation as a non-artifact of the Lattice.
On the contrary, we can offer a simple physical understanding of
this behavior, in terms of lowest Landau level dominance of the
chiral condensate. As a matter of fact, using the simple equations
of the models for the chiral condensate and for the magnetic
moment, we can show that at large magnetic field $\mu_f$ has to
saturate to one, because in this limit the higher Landau levels
are expelled from the chiral condensate; as a consequence, the
ratio of the two approaches one asymptotically.

We also obtain a saturation of the polarization within the
renormalized QM model. There are some differences, however, in
comparison with the results of the non-renormalized models. In the
former case, the asymptotic value of $\mu_f$ is charge-dependent;
moreover, the interpretation of the saturation as a lowest Landau
level (LLL) dominance is not straightforward, because the
renormalized contribution of the higher Landau levels is important
in the chiral condensate, even in the limit of very strong fields.
It is possible that the results obtained within the renormalized
model are a little bit far from true QCD. As a matter of fact, in
the renormalized model we assume that the quark self-energy is
independent on momentum; thus, when we take the limit of infinite
quark momentum in the gap equation, and absorb the ultraviolet
divergences by means of counterterms and renormalization
conditions, we implicitly assume that that quark mass at large
momenta is equal to its value at zero momentum. We know that this
is not true, see for
example~\cite{Politzer:1976tv,Langfeld:1996rn}: even in the
renormalized theory, the quark self-energy naturally cuts off the
large momenta, leading to LLL dominance in the traces of quark
propagator which are relevant for our study. Nevertheless, it is
worth to study this problem within the renormalized QM model in
its simplest version, because it helps to understand the structure
of this theory under the influence of a strong magnetic field.

In our calculations we neglect, for simplicity, the possible
condensation of $\rho-$mesons at strong
fields~\cite{Chernodub:2010qx,Chernodub:2011mc}. Vector meson
dominance~\cite{Chernodub:2010qx} and the Sakai-Sugimoto
model~\cite{Callebaut:2011uc} suggest for the condensation a
critical value of $eB_c \approx m_\rho^2\approx 0.57$ GeV$^2$,
where $m_\rho$ is the $\rho-$meson mass in the vacuum. Beside
these, a NJL-based calculation within the lowest Landau level
(LLL) approximation~\cite{Chernodub:2011mc} predicts $\rho-$meson
condensation at strong fields as well, even if in the latter case
it is hard to estimate exactly $e B_c$, mainly because of the
uncertainty of the parameters of the model. It would certainly be
interesting to address this problem within our calculations, in
which not only the LLL but also the higher Landau levels are
considered, and in which the spontaneous breaking of chiral
symmetry is kept into account self-consistently. However, this
would complicate significantly the calculational setup. Therefore,
for simplicity we leave this issue to a future project.

In~\cite{Frasca:2011zn} the computation of the polarization and of the
magnetic susceptibility has been performed both within the NJL and the Quark-Meson (QM) models;
the qualitative picture does not depend on the model considered. Moreover,
within the QM model an analytical computation of the aforementioned quantities
within the renormalized quantum effective potential is feasible (the NJL model,
at least with a contact interaction as considered in~\cite{Frasca:2011zn}, is not renormalizable).
Therefore in this review chapter we limit ourselves to summarize the results obtained
within the QM model, both numerical and analytical, deferring to the original reference
for further details.

\subsection{Non-renormalized Quark-Meson model results}
In the QM model, a meson sector described by the linear sigma
model lagrangian, is coupled to quarks via a Yukawa-type
interaction. The model is renormalizable in $D=3+1$ dimensions.
However, since we adopt the point of view of it as an effective
description of QCD, it is not necessary to use the renormalized
version of the model itself. On the contrary, it is enough to fix
an ultraviolet scale to cutoff the divergent expectation values;
the UV scale is then chosen phenomenologically, by requiring that
the numerical value of the chiral condensate in the vacuum
obtained within the model, is consistent with the results obtained
from the sum rules~\cite{Dosch:1997wb}. This is a rough
approximation of the QCD effective quark mass, which smoothly
decays at large momenta~\cite{Langfeld:1996rn,Politzer:1976tv}. In
Section IV we will use a renormalized version of the model, to
derive semi-analytically some results in the two regimes of weak
and strong fields.

The lagrangian density of the model is given by
\begin{eqnarray}
{\cal L} &=& \bar q \left[iD_\mu\gamma^\mu - g(\sigma +
i\gamma_5\bm\tau\cdot\bm\pi)\right] q
\nonumber \\
&& + \frac{1}{2}\left(\partial_\mu\sigma\right)^2 +
\frac{1}{2}\left(\partial_\mu\bm\pi\right)^2 - U(\sigma,\bm\pi)~.
\label{eq:LD1}
\end{eqnarray}
In the above equation, $q$ corresponds to a quark field in the
fundamental representation of color group $SU(3)$ and flavor group
$SU(2)$; the covariant derivative, $D_\mu = \partial_\mu - Q_f e
A_\mu$, describes the coupling to the background magnetic field,
where $Q_f$ denotes the charge of the flavor $f$. Besides,
$\sigma$, $\bm\pi$ correspond to the scalar singlet and the
pseudo-scalar iso-triplet fields, respectively. The potential $U$
describes tree-level interactions among the meson fields. In this
article, we take its analytic form as
\begin{equation}
U(\sigma,\bm\pi) = \frac{\lambda}{4}\left(\sigma^2
+\bm\pi^2-v^2\right)^2 - h\sigma~, \label{eq:U}
\end{equation}
where the first addendum is chiral invariant; the second one
describes a soft explicit breaking of chiral symmetry, and it is
thus responsible for the non-zero value of the pion mass. For
$h=0$, the interaction terms of the model are invariant under
$SU(2)_V\otimes SU(2)_A\otimes U(1)_V$. This group is broken
explicitly to $U(1)_V^3\otimes U(1)_A^3\otimes U(1)_V$ if the
magnetic field is coupled to the quarks, because of the different
electric charge of $u$ and $d$ quarks. Here, the superscript $3$
in the $V$ and $A$ groups denotes the transformations generated by
$\tau_3$, $\tau_3\gamma_5$ respectively. Therefore, the chiral
group in presence of a magnetic field is $U(1)_V^3\otimes
U(1)_A^3$. This group is then explicitly broken by the $h$-term to
$U(1)_V^3$.
 
The formalism which is used to compute the magnetic susceptibility
and the polarization of the quark condensate is similar to the one described
in the previous Sections; therefore it is not necessary to give the details 
here. It is enough to write down the expresisons for the chiral condensate
at zero temperature,
\begin{equation}
\langle\bar f f\rangle = - N_c\frac{|Q_f
eB|}{2\pi}\sum_{k=0}^\infty \beta_k \int\frac{dp_3}{2\pi}
\frac{m_q}{\omega_k(p_3) }~,\label{eq:CC}
\end{equation}
where the divergent integral on the r.h.s. of the above equation
has to be understood regularized as in~\eqref{eq:ULAMBDA}, and for the  magnetic moment for the flavor $f$,
\begin{equation}
\langle \bar f \Sigma^{\mu\nu} f\rangle =
-\text{Tr}[\Sigma^{\mu\nu}S_f(x,x)]~.\label{eq:MDef}
\end{equation}
We take $\bm B = (0,0,B)$; in
this case, only $\Sigma^{12}\equiv\Sigma_f$ is non-vanishing.
Using the properties of $\gamma-$matrices it is easy to show that
only the Lowest Landau Level (LLL) gives a non-vanishing
contribution to the trace:
\begin{equation}
\Sigma_f = N_c \frac{Q_f |eB|}{2\pi}\int\frac{dp_3}{2\pi}
\frac{m_q}{\omega_0(p_3) }~,\label{eq:TrM}
\end{equation}
where $\omega_0 = \omega_{k=0}$.

From
Eq.~\eqref{eq:CC} we notice that the prescription~\eqref{eq:ULAMBDA}
is almost equivalent to the introduction of a running effective
quark mass,
\begin{equation}
m_q = g\sigma\Theta\left(\Lambda^2 - p_3^2 - 2k|Q_f eB|\right)~,
\label{eq:RM}
\end{equation}
that can be considered as a rough approximation to the effective
running quark mass in QCD~\cite{Politzer:1976tv} which decays at
large quark momenta, see also the discussion
in~\cite{Langfeld:1996rn}. Once the scale $\Lambda$ is fixed, the
Landau levels with $n\geq 1$ are removed from the chiral
condensate if $eB \gg \Lambda^2$.

\begin{figure}[t!]
\begin{center}
\includegraphics[width=7.0cm]{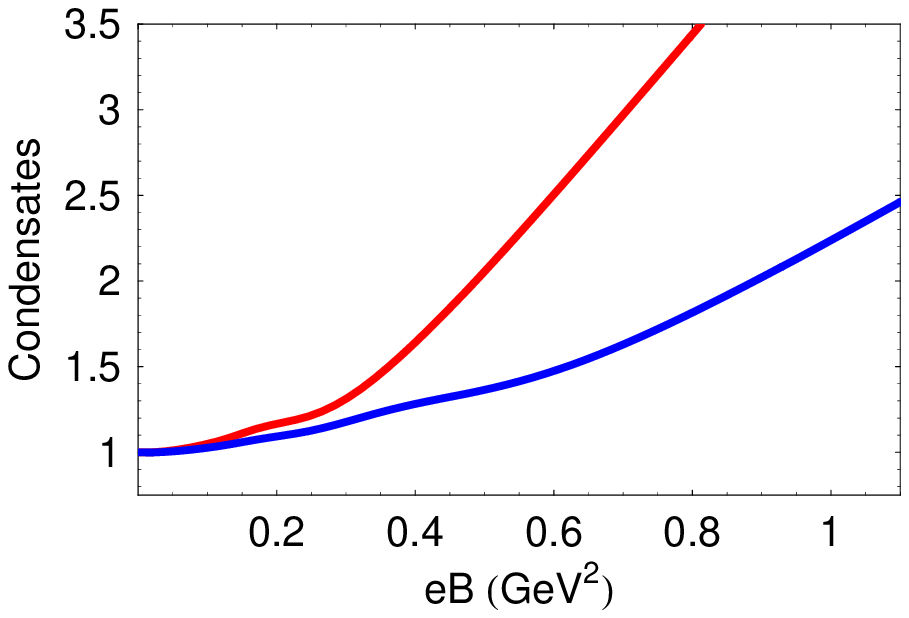}\\
\includegraphics[width=7.0cm]{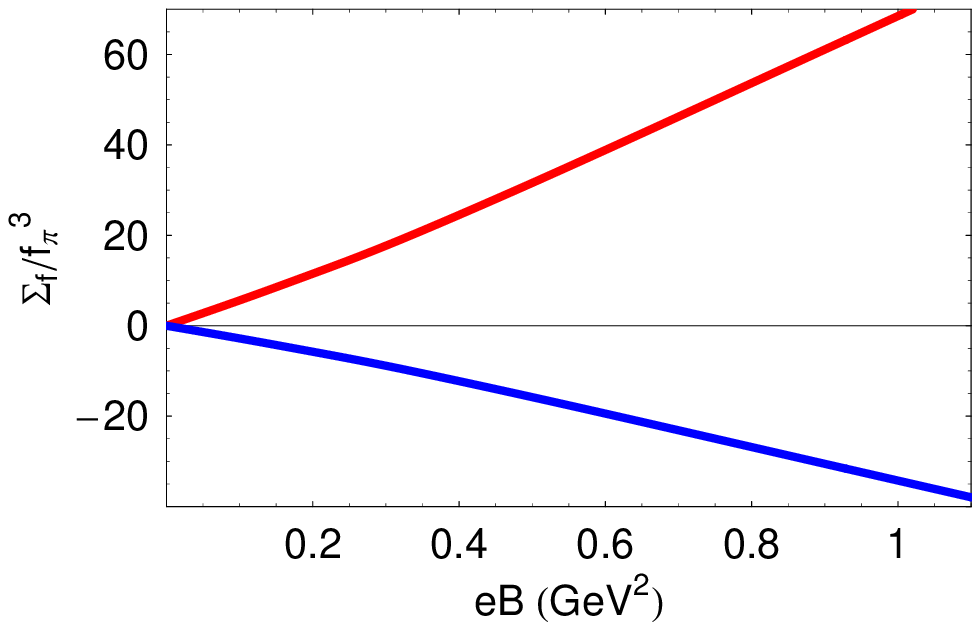}\\
\includegraphics[width=7.0cm]{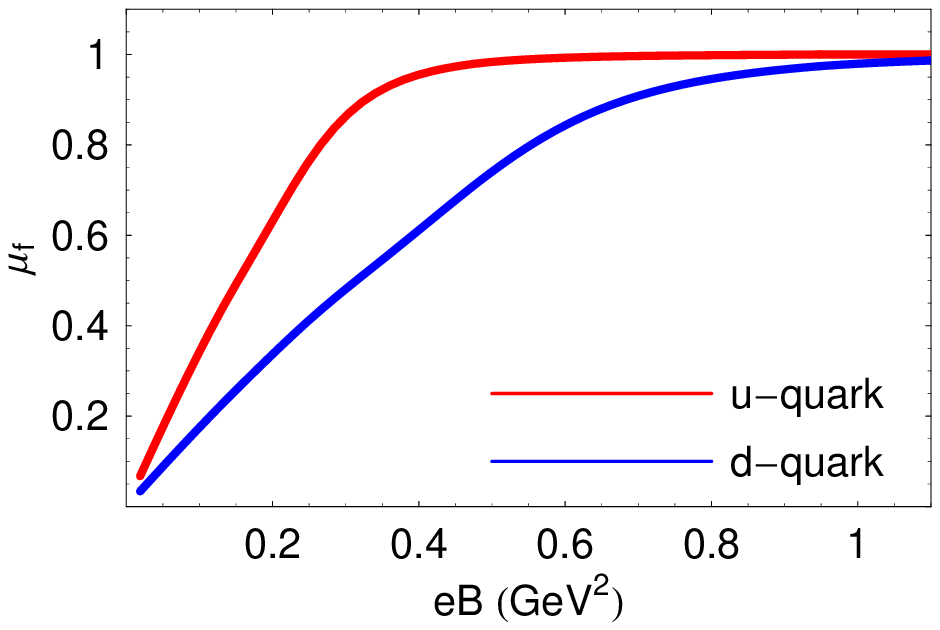}
\end{center}
\caption{\label{Fig:2} {\em Upper panel.} Chiral condensates of
$u$-quarks (red) and $d$-quarks (blue), in units of the same
quantities at zero magnetic field, as a function of the magnetic
field.  {\em Middle panel.} Expectation value of the magnetic
moment operator, in units of $f_\pi^3$, as a function of $eB$.
{\em Lower panel.} Polarization of $u$-quarks (red) and
$d-$quarks (blue) as a function the magnetic field strength. 
From Ref.~\cite{Frasca:2011zn}. Copyright(2012) by
American Physical Society.}
\end{figure}

In the upper panel of Fig.~\ref{Fig:2}, we plot the chiral
condensates for $u$ and $d$ quarks, as a function of $eB$, for the
QM model. The magnetic field splits the two quantities because of
the different charge for the two quarks. The small oscillations,
which are more evident for the case of the $u$-quark, are an
artifact of the regularization scheme, and disappear if smoother
regulators are used, see the discussion
in~\cite{Campanelli:2009sc}. In the regime of weak fields, our
data are consistent with the scaling $\langle\bar ff\rangle
\propto |eB|^2/M$ where $M$ denotes some mass scale; in the strong
field limit we find instead $\langle\bar ff\rangle \propto
|eB|^{3/2}$. The behavior of the quark condensate as a function of
magnetic field is in agreement with the magnetic catalysis
scenario.

In the middle panel of Fig.~\ref{Fig:2} we plot our data for the
expectation value of the magnetic moment. At weak
fields, $\Sigma_f \propto |eB|$ as expected from
Eq.~\eqref{eq:MDef}. In the strong field limit, non-linearity
arises because of the scaling of quark mass (or chiral condensate);
we find $\Sigma_f \propto |eB|^{3/2}$ in this limit.

In the lower panel of Fig.~\ref{Fig:2} we plot our results for the polarization. Data
are obtained by the previous ones, using the
definition~\eqref{eq:muDef}. At small fields, the polarization
clearly grows linearly with the magnetic field. This is a natural
consequence of the linear behavior of the magnetic moment as a
function of $eB$ for small fields, see Fig.~\ref{Fig:2}. On the
other hand, within the chiral models we measure a saturation of
$\mu_f$ at large values of $eB$, to an asymptotic value
$\mu_\infty = 1$. This conclusion remains unchanged if we consider
the NJL model, and it is in agreement with the recent Lattice
findings~\cite{Buividovich:2009ih}. It should be noticed that, at
least for the $u-$quark, saturation is achieved before the
expected threshold for $\rho-$meson
condensation~\cite{Chernodub:2010qx,Chernodub:2011mc,Callebaut:2011uc}.
Therefore, our expectation is that our result is stable also if
vector meson condensation is considered.

The saturation to the asymptotic value $\mu_\infty = 1$ of
polarization is naturally understood within the models we
investigate, as a LLL dominance in the chiral condensate (i.e.,
full polarization). As a matter of fact, $\Sigma_f$ and
$\langle\bar ff\rangle$ turn out to be proportional in the strong
field limit, since only the LLL gives a contribution to the the
latter, comparing Eq.~\eqref{eq:CC} and~\eqref{eq:TrM} which imply
\begin{equation}
\mu_f = 1 - \frac{\langle\bar ff\rangle_{\text{HLL}}}{\langle\bar
ff\rangle}~, \label{eq:HLL}
\end{equation}
where $\langle\bar ff\rangle_{\text{HLL}}$ corresponds to the
higher Landau levels contribution to the chiral condensate. In the
strong field limit $\langle\bar ff\rangle_{\text{HLL}} \rightarrow
0$, see Eq.~\eqref{eq:CC};
hence, $\mu_f$ has to approach the asymptotic value $\mu_\infty =
1$.  On the other hand, in the weak field limit $\langle\bar
ff\rangle_{\text{HLL}} \rightarrow \langle\bar ff\rangle$ and the
proportionality among $\Sigma_f$ and $\langle\bar ff\rangle$ is
lost.

At small fields $\mu_f = |\chi Q_f eB|$ from
Eq.~\eqref{eq:chiDef}. Hence, we use the data on polarization at
small fields, to obtain the numerical value of the magnetic
susceptibility of the chiral condensate. Our results are as
follows:
\begin{eqnarray}
\chi &\approx& -4.3~\text{GeV}^{-2}~,~~~\text{NJL}
\\
\chi &\approx& -5.25~\text{GeV}^{-2}~,~~~\text{QM}
\end{eqnarray}
respectively for the NJL model and the QM model. To obtain the
numerical values above we have used data for $eB$ up to $5 m_\pi^2
\approx 0.1$ GeV$^2$, which are then fit using a linear law. Using
the numerical values of the chiral condensate in the two models,
we obtain
\begin{eqnarray}
\chi \langle\bar f f \rangle &\approx&
69~\text{MeV}~,~~~\text{NJL} \\
\chi \langle\bar f f \rangle &\approx& 65~\text{MeV}~,~~~\text{QM}
\end{eqnarray}
The numerical values of $\chi$ that we obtain within the effective
models are in fair agreement with recent results, see Table I. In
our model calculations, the role of the renormalization scale is
played approximately by the ultraviolet cutoff, which is equal to $0.560$ GeV in the QM
model, and $0.627$ GeV in the NJL model.

To facilitate the comparison with previous estimates, we review
briefly the frameworks in which the results in Table I are
obtained. In~\cite{Vainshtein:2002nv} the following result is
found, within OPE combined with Pion Dominance (we follow when
possible the notation used in~\cite{Buividovich:2009ih}):
\begin{equation}
\chi^{PD} = -c_\chi\frac{N_c}{8\pi^2 F_\pi^2}~,~~~{\text{Pion
Dominance}} \label{eq:PD}
\end{equation}
with $F_\pi =\sqrt{2}f_\pi= 130.7$ MeV and $c_\chi = 2$; the
estimate of~\cite{Vainshtein:2002nv} is done at a renormalization
point $M = 0.5$ GeV. It is remarkable that Eq.~\eqref{eq:PD} has
been reproduced recently within AdS/QCD approach
in~\cite{Son:2010vc}. Probably, this is the result more comparable
to our estimate, because the reference scales
in~\cite{Vainshtein:2002nv} and in this article are very close.
Within our model calculations we find $c_\chi^{NJL} = 1.93$ and
$c_\chi^{QM} = 2.36$. Using the numerical value of $F_\pi$ and
$c_\chi$ we get $\chi^{PD} = -4.45$ GeV$^{-2}$, which agrees
within $3\%$ with our NJL model result, and within $18\%$ with our
QM model result.

In~\cite{Gorsky:2009ma} the authors find $c_\chi = 2.15$ within
hard-wall holographic approach, at the scale $M \ll 1$ GeV. The
results of~\cite{Gorsky:2009ma} are thus in very good parametric
agreement with~\cite{Vainshtein:2002nv}; on the other hand, the
numerical value of $F_\pi$ in the holographic model is smaller
than the one used in~\cite{Vainshtein:2002nv}, pushing the
holographic prediction for $\chi$ to slightly higher values than
in~\cite{Vainshtein:2002nv}. However, the scale at which the
result of~\cite{Gorsky:2009ma} is valid should be much smaller
than $M=1$ GeV, thus some quantitative disagreement
with~\cite{Vainshtein:2002nv} is expected. As the authors have
explained, it might be possible to tune the parameters of the
holographic model, mainly the chiral condensate, to reproduce the
correct value of $F_\pi$; their numerical tests suggest that by
changing the ratio $\langle\bar ff\rangle/m_\rho$ of a factor of
$8$, then the numerical value of $c_\chi$ is influenced only by a
$5\%$. It is therefore plausible that a best tuning makes the
quantitative prediction of~\cite{Gorsky:2009ma} much closer to the
estimate of~\cite{Vainshtein:2002nv}.

In~\cite{Kim:2004hd} an estimate of $\chi$ within the instanton
vacuum model has been performed beyond the chiral limit, both for
light and for strange quarks (the result quoted in Table I
corresponds to the light quarks; for the strange quark,
$\chi_s/\chi_{u,d} \approx 0.15$ is found). Taking into account
the numerical value of the chiral condensate in the instanton
vacuum, the numerical estimate of~\cite{Kim:2004hd} leads to $\chi
= -2.5\pm 0.15$ GeV$^{-2}$ at the scale $M=1$ GeV. An analytic
estimate within a similar framework has been obtained
in~\cite{Ioffe:2009yi}, in which the zero-mode of the Dirac
operator in the background of a $SU(2)$ instanton is used to
compute the relevant expectation values. The result
of~\cite{Ioffe:1983ju} gives $\chi = -3.52$ GeV$^{-2}$ at
$M\approx 1$ GeV.

In~\cite{Buividovich:2009ih} the result $\chi = -1.547$ GeV$^{-2}$
is achieved within a two-color simulation with quenched fermions.
It is interesting that in~\cite{Buividovich:2009ih} the same
quantity has been computed also at finite temperature in the
confinement phase, at $T=0.82 T_c$, and the result seems to be
independent on temperature. The reference scale
of~\cite{Buividovich:2009ih}, determined by the inverse lattice
spacing, is $M\approx 2$ GeV. Therefore the lattice results are
not quantitatively comparable with our model calculation. However,
they share an important feature with the results presented here,
namely the saturation of the polarization at large values of the
magnetic field. Finally, estimates of the magnetic susceptibility
of the chiral condensate by means of several QCD sum rules
exist~\cite{Ioffe:1983ju,Belyaev:1984ic,Balitsky:1985aq,Ball:2002ps,Rohrwild:2007yt}.
The results are collected in Table I.

\begin{table*}[t!]
\label{aggiungi}\centering %
\begin{tabular}{|c|c|c|c|}
\hline\hline
{\bf Method} & $\bm\chi$ (GeV$^{-2}$)& {\bf Ren. Point} (GeV)&{\bf Ref.}
\\\hline
\hline Sum rules & $-8.6\pm 0.24$ &1&~\cite{Ioffe:1983ju} \\
\hline Sum rules &-5.7&0.5&~\cite{Belyaev:1984ic} \\
\hline Sum rules & $-4.4\pm0.4$ & 1 &~\cite{Balitsky:1985aq}\\
\hline Sum rules &$-3.15\pm0.3$&1&~\cite{Ball:2002ps} \\
\hline Sum rules & $-2.85\pm0.5$ & 1 &~\cite{Rohrwild:2007yt}\\
\hline OPE + Pion Dominance & $-N_c/(4\pi^2 F_\pi^2)$  &0.5&~\cite{Vainshtein:2002nv} \\
\hline Holography & $-1.075 N_c/(4\pi^2 F_\pi^2)$  & $\ll 1$&~\cite{Gorsky:2009ma} \\
\hline Holography & $-N_c/(4\pi^2 F_\pi^2)$  & $\ll 1$&~\cite{Son:2010vc} \\
\hline Instanton vacuum &$-2.5\pm 0.15$ &1&~\cite{Kim:2004hd} \\
\hline Zero mode of Dirac Operator &-3.52&1&~\cite{Ioffe:2009yi} \\
\hline Lattice&$-1.547(3)$ &2&~\cite{Buividovich:2009ih} \\
\hline NJL model&-4.3&0.63&This work \\
\hline QM model &-5.25&0.56&This work \\
\hline
\end{tabular}
\caption{Magnetic susceptibility of the quark
condensate obtained within several theoretical approaches. In the
table, $F_\pi = 130.7$ MeV. See the text for more details. Adapted from
Ref.~\cite{Frasca:2011zn}. Copyright(2012) by American Physical Society.}
\end{table*}

\subsection{Results within the renormalized QM model}

In this Section, we make semi-analytic estimates of the
polarization and the magnetic susceptibility of the quark
condensate, as well as for the chiral condensate in magnetic
background, within the renormalized QM model. This is done with
the scope to compare the predictions of the renormalized model
with those of the effective models, in which an ultraviolet cutoff
is introduced to mimic the QCD effective quark mass.

In the renormalized model, we allow the effective quark mass to be
a constant in the whole range of momenta, which is different from
what happens in QCD~\cite{Politzer:1976tv}. Thus, the higher
Landau levels give a finite contribution to the vacuum chiral
condensate even at very strong fields. This is easy to understand:
the ultraviolet cutoff, $\Lambda$, in the renormalized model can
be taken larger than any other mass scale, in particular $\Lambda
\gg |eB|^{1/2}$; as a consequence, the condition $p_3^2 + 2 n |eB|
< \Lambda^2$ is satisfied taking into account many Landau levels
even at very large $eB$. The contribution of the higher Landau
levels, once renormalized, appears in the physical quantities to
which we are interested here, in particular in the chiral
condensate.

Since the computation is a little bit lengthy, it is useful to
anticipate its several steps: firstly we perform regularization,
and then renormalization, of the QEP at zero magnetic field (the
corrections due to the magnetic field turn out to be free of
ultraviolet divergences). Secondly, we solve analytically the gap
equation for the $\sigma$ condensate in the limit of weak fields,
and semi-analytically in the opposite limit. The field-induced
corrections to the QEP and to the solution of the gap equation are
divergence-free in agreement with~\cite{Suganuma:1990nn}, and are
therefore independent on the renormalization scheme adopted. Then,
we compute the renormalized and self-consistent values of the
chiral condensate and of the magnetic moment, as a function of
$eB$, using the results for the gap equation. Within this
theoretical framework, it is much more convenient to compute
$\langle\bar ff\rangle$ and $\Sigma_f$ by taking derivatives of
the renormalized potential; in fact, the computation of the traces
of the propagator in the renormalized model is much more involved
if compared to the situation of the non-renormalized models, since
in the former a non-perturbative (and non-trivial) renormalization
procedure of composite local operators is
required~\cite{Collins:1984xc}. Finally, we estimate $\chi$, as
well as the behavior of the polarization as a function of $eB$.


\subsubsection{Renormalization of the QEP}

To begin with, we need to regularize the one-loop fermion
contribution, namely
\begin{eqnarray}
V_{1-\text{loop}}^\text{fermion} &=& - N_c\sum_f\frac{|Q_f
eB|}{2\pi}
\sum_{n=0}^\infty\beta_n
\int_{-\infty}^{+\infty}\frac{dk}{2\pi} \left(k^2 + 2n|Q_f eB| +
m_q^2\right)^{1/2}~.
\end{eqnarray}
To this end, we define the function, ${\cal V}(s)$, of a complex
variable, $s$, as
\begin{eqnarray}
{\cal V}(s) &=& -N_c\sum_f\frac{|Q_f
eB|}{2\pi}
\sum_{n=0}^\infty\beta_n
\int_{-\infty}^{+\infty}\frac{dk}{2\pi} \left(k^2 + 2n|Q_f eB| +
m_q^2\right)^\frac{1-s}{2}~. 
\label{eq:Vfs}
\end{eqnarray}
The function ${\cal V}(s)$ can be analytically continued to $s=0$.
We define then $V^{\text{fermion}}_{1-\text{loop}} =
\lim_{s\rightarrow 0^+} {\cal V}(s)$. After elementary integration
over $k$, summation over $n$ and taking the limit $s\rightarrow
0^+$, we obtain the result
\begin{eqnarray}
V_{1-\text{loop}}^\text{fermion} &=& N_c\sum_f\frac{(Q_f
eB)^2}{4\pi^2}\left(\frac{2}{s} -
\log(2|Q_f eB|)+a\right) B_2(q) \nonumber\\
&&-
N_c\sum_f\frac{(Q_f eB)^2}{2\pi^2}\zeta^\prime\left(-1,q\right) \nonumber\\
&&-N_c\sum_f\frac{|Q_f eB| m_q^2}{8\pi^2}\left(\frac{2}{s} -
\log(m_q^2)+a\right)~, \label{eq:F4}
\end{eqnarray}
where we have subtracted terms which do not depend explicitly on
the condensate. In the above equation, $\zeta\left(t,q\right)$ is
the Hurwitz zeta function; for $\text{Re}(t) > 1$ and
$\text{Re}(q) > 0$, it is defined by the series
$\zeta\left(t,q\right) = \sum_{n=0}^\infty(n+q)^{-t}$; the series
can be analytically continued to a meromorphic function defined in
the complex plane $t \neq 1$. Moreover we have defined $q = (m_q^2
+ 2|Q_f eB|)/2|Q_f eB|$; furthermore, $a = 1 - \gamma_E -
\psi(-1/2)$, where $\gamma_E$ is the Eulero-Mascheroni number and
$\psi$ is the digamma function. The derivative $\zeta^\prime
\left(-1,q\right) =d\zeta(t,q)/d t$ is understood to be computed
at $t=-1$.

The first two addenda in Eq.~\eqref{eq:F4} arise from the higher
Landau levels; on the other hand, the last addendum is the
contribution of the LLL. The function $B_2$ is the second
Bernoulli polynomial; using its explicit form, it is easy to show
that the divergence in the LLL term in Eq.~\eqref{eq:F4} is
canceled by the analogous divergence in the first addendum of the
same equation. It is interesting that the LLL contribution, which
is in principle divergent, combines with a part of the
contribution of the higher Landau levels, leading to a finite
result. This can be interpreted as a renormalization of the LLL
contribution. On the other hand, the remaining part arising from
the higher Landau levels is still divergent; this divergence
survives in the $\bm B\rightarrow 0$ limit, and is due to the
usual divergence of the vacuum contribution. We then have
\begin{eqnarray}
V^{\text{fermion}}_{1-\text{loop}} &=&
N_c\sum_f\frac{m_q^4}{16\pi^2}\left(\frac{2}{s}-\log(2|Q_f eB|)+a\right)\nonumber\\
&&+N_c\sum_f\frac{|Q_f eB| m_q^2}{8\pi^2}\log\frac{m_q^2}{2|Q_f
eB|} \nonumber
\\
&&-N_c\sum_f \frac{(Q_f
eB)^2}{2\pi^2}\zeta^\prime\left(-1,q\right)~. \label{eq:F4bis}
\end{eqnarray}
The renormalization procedure of the quantum effective potential is
discussed in some detail in~\cite{Frasca:2011zn}. Here it is not necessary
to discuss this procedure, and we just focus on the results.

\subsubsection{Approximate solutions of the gap equation}
{\em Weak fields.} In the weak field limit ($eB
\ll m_q^2$) the correction due to the magnetic field to the quantum effective potential
can be computed:
\begin{eqnarray}
V_1&\approx& - N_c\sum_f\frac{(Q_f
eB)^2}{24\pi^2}\log\frac{m_q^2}{2|Q_f e B|} 
=- N_c\sum_f\frac{(Q_f eB)^2}{24\pi^2}\log\frac{m_q^2}{\mu^2}
~, \label{eq:cotta}
\end{eqnarray}
which is in agreement with the result of~\cite{Suganuma:1990nn}.
In the above equation we have followed the notation
of~\cite{Cohen:2008bk} introducing an infrared scale $\mu$,
isolating and then subtracting the term which does not depend on
the condensate. The scale $\mu$ is arbitrary, and we cannot
determine it from first principles; on the other hand, it is
irrelevant for the determination of the $\sigma-$condensate. We
expect $\mu\approx f_\pi$ since this is the typical scale of
chiral symmetry breaking in the model for the $\sigma$ field. 

In this limit, it is easy to obtain analytically the behavior of
the constituent quark mass as a function of $eB$. As a matter of
fact, we can expand the derivative of the QEP with respect to
$\sigma$, around the solution at $B=0$, writing
$\langle\sigma\rangle = f_\pi + \delta\sigma$. Then, a
straightforward evaluation leads to
\begin{equation}
m_q = g f_\pi\left( 1 + \frac{5}{9}\frac{N_c}{12\pi^2 f_\pi^2
m_\sigma^2}  (eB)^2 \right)~. \label{eq:WEA}
\end{equation}
As anticipated, the scale $\mu$ is absent in the solution of the
gap equation.

{\em Strong fields.} In the limit $eB\gg m_q^2$, we can find an
asymptotic representation of $V_1$ by using the expansion
$\zeta^\prime(-1,q) = c_0 + c_1 (q-1)$ valid for $q\approx 1$,
with $c_0 = -0.17$ and $c_1 = -0.42$. Then we find
\begin{eqnarray}
V_1 &\approx& -N_c\sum_f \frac{m^2_q}{8\pi^2}\left(\frac{m^2_q}{2}
+ |Q_f eB|\right)\log\frac{2|Q_f eB|}{m^2_q} 
- N_c\sum_f\frac{|Q_f eB|m_q^2}{2\pi^2}c_1~, \label{eq:Ft}
\end{eqnarray}
where we have subtracted condensate-independent terms.

In the strong field limit it is not easy to find analytically an
asymptotic representation for the sigma condensate as a function
of $eB$; therefore we solve the gap equation numerically, and then
fit data with a convenient analytic form as follows:
\begin{equation}
m_q = b|eB|^{1/2} + \frac{c f_\pi^3}{|eB|}~, \label{eq:SEA}
\end{equation}
where $b=0.32$ and $c=32.78$. At large fields the quark mass grows
as $|eB|^{1/2}$ as expected by dimensional analysis; this is a
check of the equations that we use.

\subsubsection{Evaluation of chiral condensate and magnetic moment}
{\em Chiral condensate.} To compute the chiral condensate we
follow a standard procedure: we introduce source term for $\bar
ff$, namely a bare quark mass $m_f$, then take derivative of the
effective potential with respect to $m_f$ evaluated at $m_f = 0$.
For the weak field case we obtain
\begin{eqnarray}
\langle\bar ff\rangle &=& \langle\bar ff\rangle_0 -
\frac{N_c}{12\pi^2}\frac{ |Q_f e B|^2}{m_q}~. \label{eq:CCggg}
\end{eqnarray}
On the other hand, in the strong field limit we have
\begin{equation}
\langle\bar ff\rangle = -\frac{N_c m_q}{4\pi^2} \left(|Q_f e B| +
m_q^2\right)\log\frac{2|Q_f e B|}{m_q^2} ~. \label{eq:CCsss}
\end{equation}
Using Equations~\eqref{eq:WEA} and~\eqref{eq:SEA}, we show that
the chiral condensate scales as $a + b (eB)^2$ for small fields,
and as $|eB|^{3/2}$ for large fields.

{\em Magnetic moment.} Next we turn to the computation of the
expectation value of the magnetic moment. The expression in terms
of Landau levels is given by Eq.~\eqref{eq:TrM}, which clearly
shows that this quantity has a log-type divergence. In order to
avoid a complicated renormalization procedure of a local composite
operator, we notice that it is enough to take the minus derivative
of $V_1$ with respect to $\bm B$ to get magnetization, ${\cal
M}$~\cite{Cohen:2008bk}, then multiply by $2m/Q_f$ to get the
magnetic moment. This procedure is very cheap, since the $\bm
B-$dependent contributions to the effective potential are finite,
and the resulting expectation value will turn out to be finite as
well (that is, already renormalized).

In the case of weak fields, from Eq.~\eqref{eq:cotta} we find
\begin{equation}
\Sigma_f = N_c \frac{Q_f |eB|
m_q}{6\pi^2}\log\frac{m_q^2}{\mu^2}~. \label{eq:WFmm}
\end{equation}
On the other hand, in the strong field limit we get from
Eq.~\eqref{eq:Ft}
\begin{equation}
\Sigma_f = N_c \frac{m_q^3}{4\pi^2}\log\frac{2|Q_f eB|}{m_q^2}~.
\label{eq:SFmm}
\end{equation}
The above result is in parametric agreement with the estimate of
magnetization in~\cite{Cohen:2008bk}. In fact, $m_q^2\approx |eB|$
in the strong field limit, which leads to a magnetization ${\cal
M}\approx B\log B$.

Using the expansions for the sigma condensate at small and large
values of the magnetic field strength, we argue that $\Sigma_f
\approx |eB|$ in a weak field, and $\Sigma_f \approx |eB|^{3/2}$
in a strong field.

\subsubsection{Computation of chiral magnetization and polarization}
We can now estimate the magnetic susceptibility of the quark
condensate and the polarization as a function of $eB$. For the
former, we need to know the behavior of the magnetic moment for
weak fields. From Eq.~\eqref{eq:WFmm} and from the
definition~\eqref{eq:chiDef} we read
\begin{equation}
\chi\langle\bar ff\rangle =\frac{N_c
m_q}{6\pi^2}\log\frac{m_q^2}{\mu^2}\equiv f(\mu)~.
\label{eq:slope}
\end{equation}
The presence of the infrared scale $\mu$ makes the numerical
estimate of $\chi$ uncertain; however, taking for it a value
$\mu\approx f_\pi$, which is the typical scale of chiral symmetry
breaking, we have $\chi\langle\bar ff\rangle \approx 44$ MeV,
which is in agreement with the expected value, see
Eq.~\eqref{eq:i1}. 



Next we turn to the polarization. For weak fields we find
trivially a linear dependence of $\mu_f$ on $|Q_f eB|$, with slope
given by the absolute value of $\chi$ in Eq.~\eqref{eq:slope}. On
the other hand, in the strong field limit we find, according to
Eq.~\eqref{eq:SFmm},
\begin{equation}
\mu_f \approx \frac{m_q^2}{m_q^2 + |Q_f eB|} \approx 1 -
\frac{|Q_f|}{b + |Q_f|}~, \label{eq:asymSF}
\end{equation}
where we have used Eq.~\eqref{eq:SEA}. This result shows that the
polarization saturates at large values of $eB$, but the asymptotic
value depends on the flavor charge.

It is interesting to compare the result of the renormalized model
with that of the effective models considered in the previous
Section. In the former, the asymptotic value of $\mu_f$ is
flavor-dependent; in the latter, $\mu_f \rightarrow 1$
independently on the value of the electric charge. Our
interpretation of this difference is as follows: comparing
Eq.~\eqref{eq:asymSF} with the general model expectation,
Eq.~\eqref{eq:HLL}, we recognize in the factor $|Q_f|/(b+|Q_f|)$
the contribution of the higher Landau levels at zero temperature,
which turns out to be finite and non-zero after the
renormalization procedure. This contribution is then transmitted
to the physical quantities that we have computed. The trace of the
higher Landau levels is implicit in the solution of the gap
equation in the strong field limit, namely the factor $b$ in
Eq.~\eqref{eq:SEA}, and explicit in the additional $|Q_f|$
dependence in Eq.~\eqref{eq:asymSF}. A posteriori, this conclusion
seems quite natural, because in the renormalization procedure we
assume that the effective quark mass is independent of quark
momentum, thus there is no cut of the large momenta in the gap
equation (and in the equation for polarization as well). In the
effective models considered in the first part of this article, on
the other hand, the cutoff procedure is equivalent to have a
momentum-dependent effective quark mass, $m_q =
g\sigma\Theta(\Lambda^2 - p_3^2 - 2n|Q_f eB|)$, which naturally
cuts off higher Landau levels when $eB \gg \Lambda^2$. At the end
of the days, the expulsion of the higher Landau levels from the
chiral condensate makes $\mu_f \rightarrow 1$ in the strong field
limit. Our expectation is that if we allow the quark mass to run
with momentum and decay rapidly at large momenta, mimicking the
effective quark mass of QCD, higher Landau levels would be
suppressed in the strong field limit, and the
result~\eqref{eq:asymSF} would tend to the result in
Fig.~\ref{Fig:2}.

\section{Conclusions}
In this Chapter we have summarized our results for the phase structure of quark matter
in a strong magnetic background. Our theoretical investigative tools are 
chiral quark models improved with the Polyakov loop, which allow to study simultaneously chiral symmetry breaking
and deconfinement.  

The main motivation of this series of studies is of a phenomenological nature. 
In fact, it has been shown that huge magnetic fields are produced in non-central
heavy ion collisions. These fields might trigger the $P-$ and $CP-$odd
process dubbed Chiral Magnetic Effect (CME). Therefore, in order to make quantitative
estimates of the observables which are sensitive to the CME, it is extremely important to understand
how hot quark matter behaves under the influence of a strong magnetic background.
The other side of the coin is that simulations show these huge fields have
a very short lifetime: therefore the present studies should take into account
this time dependence. Moreover electric fields, which we have neglected so far, 
are also produced in the collisions. Finally, the electromagnetic fields considerably depend
on space coordinates on the scale of the volume of the expanding fireball. This dependence
has been ignored in our studies, since the magnetic background is taken to be homogeneous
in space and constant in time. 

Our results support the scenario of magnetic catalysis, which manifests itself in both
an increase of the chiral condensate at zero temperature, and an increase of the critical temperature
for chiral symmetry restoration. Moreover, depending on the interaction used,
deconfinement may occur either together with chiral symmetry restoration,
or anticipate it. The latter possibility, even if more fascinating than the former
since it opens a window for the Constituent Quark Phase, seems to be excluded
by lattice simulations.

Recent lattice simulations show that the critical temperature for chiral symmetry restoration, $T_c$,
is strongly affected by the quark mass. In particular, for small quark masses (hence, for the $u$
and $d$ quarks) the critical temperature {\em decreases} with the magnetic field strength; on the other
hand,  $T_c$ increases with the magnetic field strength for the $s$ quark. As we have discussed
in the main body of this Chapter, it seems that self-consistent computations within chiral
quark models are not able to reproduce this feature, even when quantum fluctuations are
taken into account. Thus, it remains an open problem to understand this unexpected behavior
of $T_c$. Certainly independent simulations performed by other groups are necessary to confirm
the present results.

We have also briefly summarized a computation of the magnetic susceptibility of the chiral condensate,
$\chi$, and of quark polarization, $\mu_f$ at zero temperature, based on the quark-meson model. The computed value
of $\chi$ is in agreement with most of the previous estimates, and with experimental data. 
Moreover, this model gives a simple
interpretation of the saturation of $\mu_f$ observed on the lattice: at very large magnetic field
strength, the quarks occupy the lowest Landau level, expelling higher levels from the chiral
condensate; hence, chiral condensate turns out to be proportional to the quark magnetic moment,
making the ratio (that is, polarization) just a constant. In the case of the non-renormalized model,
this constant turns out to be equal to one and flavor independent; 
on the other hand, in the case of the renormalized model, the constant is flavor dependent.
The latter result is easily understood: the renormalization procedure of the momentum indepentent interaction
of the quark-meson model brings all the Landau levels into the renormalized chiral condensate.
We expect that the replacement of the simple interaction discussed here with a non-local
one, which should mimick the quark self-energy measured on the lattice, will make the
expulsion of the higher Landau levels active also in the renormalized model, hence reproducing
the results of the non-renormalized model.

\begin{acknowledgement}
We are pleased to acknowledge the editors of this volume of {\em Lecture Notes in Physics}
for their interest in our work and their kind invitation to contribute to the book.
We also acknowledge K.~Fukushima and M.~Frasca for scientific collaborations which led to some of the
results presented here. 
\end{acknowledgement}

\end{document}